\documentclass{aa}
\pdfoutput=1 
\usepackage{graphicx}
\usepackage{txfonts}

\newcommand{\cotre}{CO (3--2)}
\newcommand{\cocinque}{CO (5--4)}
\newcommand{\cosei}{CO (6--5)}

\newcommand{\conove}{CO (9--8)}
\newcommand{\coundici}{CO (11--10)}
\newcommand{\coquattordoci}{CO (14--13)}
\newcommand{\cosedici}{CO (16--15)}
\newcommand{\codicotto}{CO (18--17)}
\newcommand{\coventi}{CO (20--19)}
\newcommand{\coventidue}{CO (22--21)}
\newcommand{\trecocinque}{$^{13}$CO (5--4)}
\newcommand{\watcsn}{H$_2$O (2$_{21}$--1$_{10}$)}
\newcommand{\watco} {H$_2$O (2$_{12}$--1$_{01}$)}
\newcommand{\oi}{$[\ion{O}{i}]$}

\newcommand{\wmq}{W m$^{-2}$}
\newcommand{\um}{$\mu$m}
\newcommand{\kms}{km~s\ensuremath{^{-1}}}
\newcommand{\cmdue}{cm$^{-2}$}
\newcommand{\cmtre}{cm$^{-3}$}
\newcommand{\unitmap}{erg s$^{-1}$ cm$^{-2}$ sr$^{-1}$}
\newcommand{\lsun}{L$_{\odot}$}

\newcommand{\jup}{$J_{\rm up}$}

\def \arcsec{\hbox{$^{\prime\prime}$}}
\def \arcmin{\hbox{$^{\prime}$}}

\begin{document}

\title{The shocked gas of the BHR71 outflow observed by Herschel\thanks{{\it Herschel} is an ESA space observatory with science instruments provided by European-led Principal Investigator consortia and with important participation from NASA.}: indirect evidence for an atomic jet }
\titlerunning{The BHR 71 outflow observed by {\it Herschel}}

\author{M. Benedettini\inst{1}, A. Gusdorf\inst{2,3}, B. Nisini\inst{4}, B. Lefloch\inst{5}, S. Anderl\inst{5,6}, G. Busquet\inst{7}, C. Ceccarelli\inst{5}, C. Codella\inst{8}, S. Leurini\inst{9}, L. Podio\inst{8}}

\institute{
INAF, Istituto di Astrofisica e Planetologia Spaziali, via Fosso del Cavaliere 100, 00133 Roma, Italy \\
\email{milena.benedettini@iaps.inaf.it}
\and
LERMA, Observatoire de Paris, PSL Research University, CNRS, UMR 8112, 75014, Paris, France
\and
Sorbonne Universit\'{e}s, UPMC Univ. Paris 6, UMR 8112, LERMA, 75005, Paris, France
\and 
INAF, Osservatorio Astronomico di Roma, via di Frascati 33, 00040, Monte Porzio Catone, Italy 
\and 
Univ. Grenoble Alpes, IPAG, F-38000 Grenoble, France
\and 
CNRS, IPAG, F-38000 Grenoble, France
\and
Institut de Ci\`{e}ncies de l’Espai (IEEC-CSIC), Campus UAB, Carrer de can Magrans, s/n, E-08193, Barcellona, Catalunya, Spain
\and
INAF, Osservatorio Astonomico di Arcetri, Largo E. Fermi 5, 50125, Firenze, Italy 
\and
Max-Planck-Institut f\"{u}r Radioastronomie, Auf dem H\"{u}gel 69, 53121, Bonn, Germany          }
\authorrunning{Benedettini et al.}

   \date{Received; accepted}

 
  \abstract
   {In the BHR71 region, two low-mass protostars IRS1 and IRS2 drive two distinguishable outflows. They constitute an ideal laboratory to investigate both the effects of shock chemistry and the mechanisms that led to their formation.} 
   {We aim to define the global morphology of the warm gas component of the BHR 71 outflow and at modelling its shocked component. }
   {We present the first far infrared {\it Herschel} images of the BHR71 outflows system in the \coquattordoci, \watcsn, \watco\, and \oi\, 145 \um\,transitions, revealing the presence of several knots of warm, shocked gas associated with the fast outflowing gas. In two of these knots we performed a detailed study of the physical conditions by comparing a large set of transitions from several molecules to a grid of shock models.}
   {The {\it Herschel} lines ratios in the outflow knots are quite similar, showing that the excitation conditions of the fast moving gas do not change significantly within the first $\sim$ 0.068 pc of the outflow, apart at the extremity of the southern blue-shifted lobe that is expanding outside the parental molecular cloud. Rotational diagram, spectral line profile and LVG analysis of the CO lines in knot A show the presence of two gas components: one extended, cold ($T\sim$80~K) and dense ($n$(H$_2$) = 3$\times$10$^5$ -- 4$\times$10$^6$~\cmtre) and another compact (18\arcsec), warm ($T$ = 1700 -- 2200~K) with slightly lower density ($n$(H$_2$) = 2$\times$10$^4$ -- 6$\times$10$^4$~\cmtre). In the two brightest knots (where we performed shock modelling) we found that H$_2$ and CO are well fitted with non-stationary (young) shocks. These models, however, significantly underestimate the observed fluxes of \oi\, and OH lines, but are not too far off those of H$_2$O, calling for an additional, possibly dissociative, J-type shock component. }
   {Our modelling indirectly suggests that an additional shock component exists, possibly a remnant of the primary jet. Direct, observational evidence for such a jet must be searched for.}


   \maketitle
%

\section{Introduction}

  \begin{figure*}
   \centering
   \includegraphics[width=16cm]{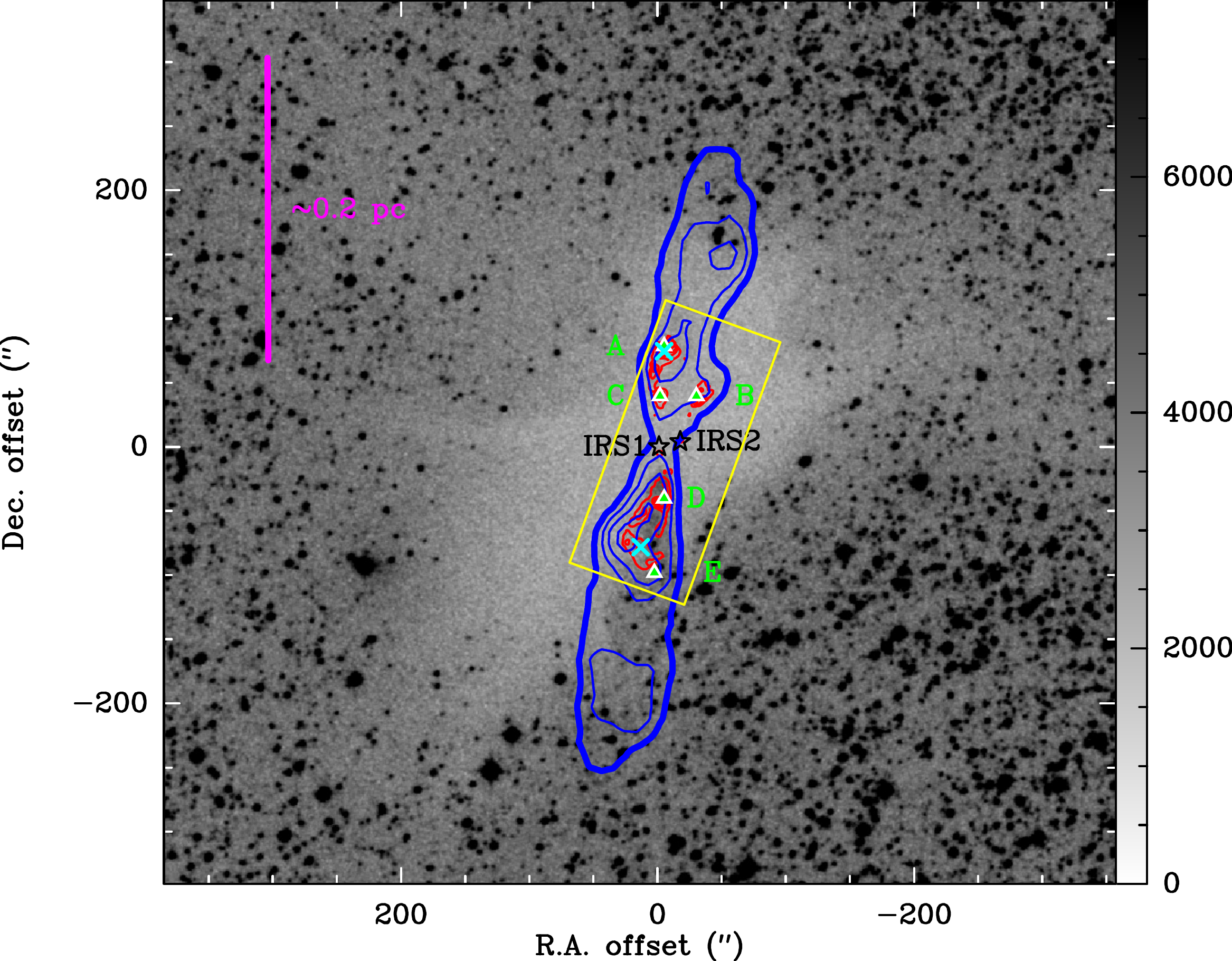}
   \caption{Double outflow of BHR71 breaking out from its parental Bok globule. The greyscale image is from the DSS2-red acquired with the UK Schmidt Telescope (\textcopyright 1993-1995 by the Anglo-Australian Telescope Board). Blue and red contours are respectively CO (3--2) and H$_2$ 0--0 S(5) emission \citep{gusdorf15}. Black stars indicate the position of the two protostars - IRS1 and IRS2 - that drive the outflows. The white and green triangles indicate the position of the peaks of the {\it Herschel} line emission and are labelled from A to E in order of decreasing declination. Yellow rectangle shows the approximate extent of the large-scale {\it Herschel}-PACS maps.The two cyan crosses indicate the positions observed with additional pointed {\it Herschel} observations.}
   \label{fig_full_sistem}
   \end{figure*}

In the early protostellar stages, fast jets powered by the nascent star, possibly surrounded by a wider angle wind, strongly interact with the parental medium through molecular shocks, producing a slower moving molecular outflow cavity. Several episodes of bow shocks are usually present within the outflow lobes \citep[e.g.][]{bachiller97,bourke97,lee06}. In several outflows, fast and well collimated molecular bullets, tracing internal shocks within the jet are also seen \citep[e.g.][]{gueth99,nisini07,tafalla10}. Outflow shocks significantly affect the gas chemical composition, with abundance enhancements up to several orders of magnitude for various molecular species such as SiO, CH$_3$OH, CS, H$_2$CO, and H$_2$O \citep[e.g.][]{bachiller01,garay98,garay02,benedettini07,arce07,codella10}. Such enrichment reveals the activation of endothermic reactions by shock heating and the mechanical erosion and processing of grains. Therefore, outflow shocks also offer a unique laboratory for testing astrochemical models. However, up to now  enough observations to allow detailed chemical studies that include multiple molecular species have been made in only a handful of outflows.

One interesting target for studying shocks associated with outflows is the BHR71 system. Located at a distance of about 175 pc \citep{bourke95}, the Bok globule BHR71 is a well-known example of an isolated star formation with a clear bipolar outflow close to the plane of the sky observable from the southern hemisphere (see Fig. \ref{fig_full_sistem}). The region comprises two distinct outflows \citep{bourke97,bourke01,parise06} powered by two protostellar sources, IRS1 and IRS2 which have luminosities of 13.5 and 0.5 \lsun, respectively \citep{chen08} and are separated by $\sim$3400 AU (assuming a distance of 200 pc, \citealt{bourke01}). A first view of the global energetics and physical conditions of the outflows have been derived by low \jup\, CO and H$_2$ observations \citep{neufeld09,giannini11}. Bright HH objects, HH320 and HH321 \citep{corporon97}, and several other knots of shocked gas have been found in the outflows. These knots aroused a great deal of interest, and their shock conditions have been studied in several papers (\citealt{giannini04}; \citealt{gusdorf11}, 2015, hereafter \citealt{gusdorf15}). The two HH objects have been imaged also in the \ion{S}{ii} transition at 6711 \AA{}, indicating that at least part of the outflows are dissociative \citep{corporon97}. In fact, suggestions of the presence of an atomic jet have been recently found by \oi\, observations \citep{nisini15}.

In this paper we present the first far infrared images of the BHR71 outflows system observed with {\it Herschel}-PACS, mapping the region of 200\arcsec\, centred on the powering sources, plus some pointed spectra of CO transitions observed with {\it Herschel}-HIFI. The images trace the not-yet-studied warm gas, at intermediate excitation conditions with respect the colder gas traced by low \jup\, CO lines \citep{parise06,gusdorf15} and the hot gas traced by H$_2$ rotational lines \citep{neufeld09,giannini11}. The main aim of this work is to define the global morphology of the warm gas component and to derive general indication on its physical condition. Moreover, in the two brightest infrared knots we performed a detailed study of the shock conditions by fitting a large set of molecular transitions from different species with shock models. The paper is organised as follows. In Sect. 2 we present the new data acquired with {\it Herschel} and the reduction method. The results and the analysis performed on the {\it Herschel} data are described in Sects. 3 and 4. In Sect. 5 we present the deeper analysis of the large set of CO transitions available for the brightest knot of the northern lobe. Shock models for this knot an another knot of the southern lobe are presented in Sect. 6. The final conclusions are reported in Sect. 7.

\section{Observations and data reduction}

We present observations carried out towards the BHR71 outflows system with the PACS  \citep{poglitsch10} and HIFI \citep{degraauw10} instruments on-board {\it Herschel} \citep{pilbratt10}. For the scientific analysis we also make use of already published {\it Herschel} and APEX observations. In particular, the map in the [\ion{O}{i}] $^3P_1$--$^3P_2$ line \citep{nisini15} and in the \cosei\, line \citep{gusdorf15}. A summary of the new {\it Herschel} observations is presented in Table~\ref{obs_lines}.

\begin{table*}
\caption{Observations towards the BHR71 outflow.}             
\label{obs_lines}      
\centering 

\begin{tabular}{c c c c c c c c}   

\multicolumn{8}{c}{220\arcsec$\times$100\arcsec, spatially Nyquist-sampled maps} \\

\hline\hline                 
Species & transition & $E_{\rm up}/k_B$ & Frequency & Wavelength & Instrument & FWHM    & Spectral resolution \\
        &            & (K)              & (GHz)     &  ($\mu$m ) &            & (\arcsec)& (\kms) \\    
\hline                        
H$_2$O & (2$_{21}$--1$_{10}$)& 159.8 & 2773.977 & 108.07 & PACS &  9.4  & 318 \\
$[\ion{O}{i}]$ & $^3P_0$--$^3P_1$ & 326.6 & 2060.069& 145.53 & PACS & 10.2  & 260 \\
H$_2$O & (2$_{12}$--1$_{01}$)& 80.1 & 1669.905 & 179.53 & PACS & 13.2  & 210 \\
CO     & (14--13)            & 580.5 &1611.794 & 186.00 & PACS & 13.2  & 195 \\

\hline

 \multicolumn{8}{c}{47\arcsec$\times$47\arcsec\, at 9\farcs4 spatial sampling maps} \\
  
\hline\hline                 
CO  & (22--21) & 1397.3 & 2528.172 & 118.58 & PACS & 9.4   & 300 \\
OH  &$^2\Pi_{\frac{3}{2}}-^2\Pi_{\frac{3}{2}} J={\frac{5}{2}}^--{\frac{3}{2}}^+$ & 120.7 & 2514.298 & 119.23 & PACS& 9.4   & 300 \\
OH  &$^2\Pi_{\frac{3}{2}}-^2\Pi_{\frac{3}{2}} J={\frac{5}{2}}^+-{\frac{3}{2}}^-$ & 120.7 & 2509.934 & 119.44 & PACS& 9.4   & 300 \\
CO  & (20--19) & 1109.2 & 2299.570 & 130.37 & PACS & 9.4   & 275  \\
CO  & (18--17) & 944.9 & 2070.616 & 144.78 & PACS & 10.2  & 260  \\
$[\ion{O}{i}]$ & $^3P_0$--$^3P_1$ & 326.6 & 2060.069& 145.53 & PACS & 10.2  & 260 \\
CO  & (14--13) & 580.5 & 1611.794 & 186.00 & PACS & 13.2  & 195  \\
\hline

 \multicolumn{8}{c}{Single spectra} \\
 
\hline\hline                 
CO        & (14--13) & 580.5  & 1611.794 & 186.00 & HIFI & 13.2 & 0.093 \\
CO        &  (9--8)  & 248.9  & 1036.912 & 289.12 & HIFI & 21.7 & 0.14 \\CO        &  (5--4)  & 83.0   &  576.268 & 520.23 & HIFI & 39.0 & 0.26 \\$^{13}$CO &  (5--4)  & 79.3   &  550.926 & 540.73 & HIFI & 40.8 & 0.27 \\\hline                       
\end{tabular}
\end{table*}

\subsection{{\it Herschel} line maps}

\begin{figure*}
   \centering
   \includegraphics[width=16cm]{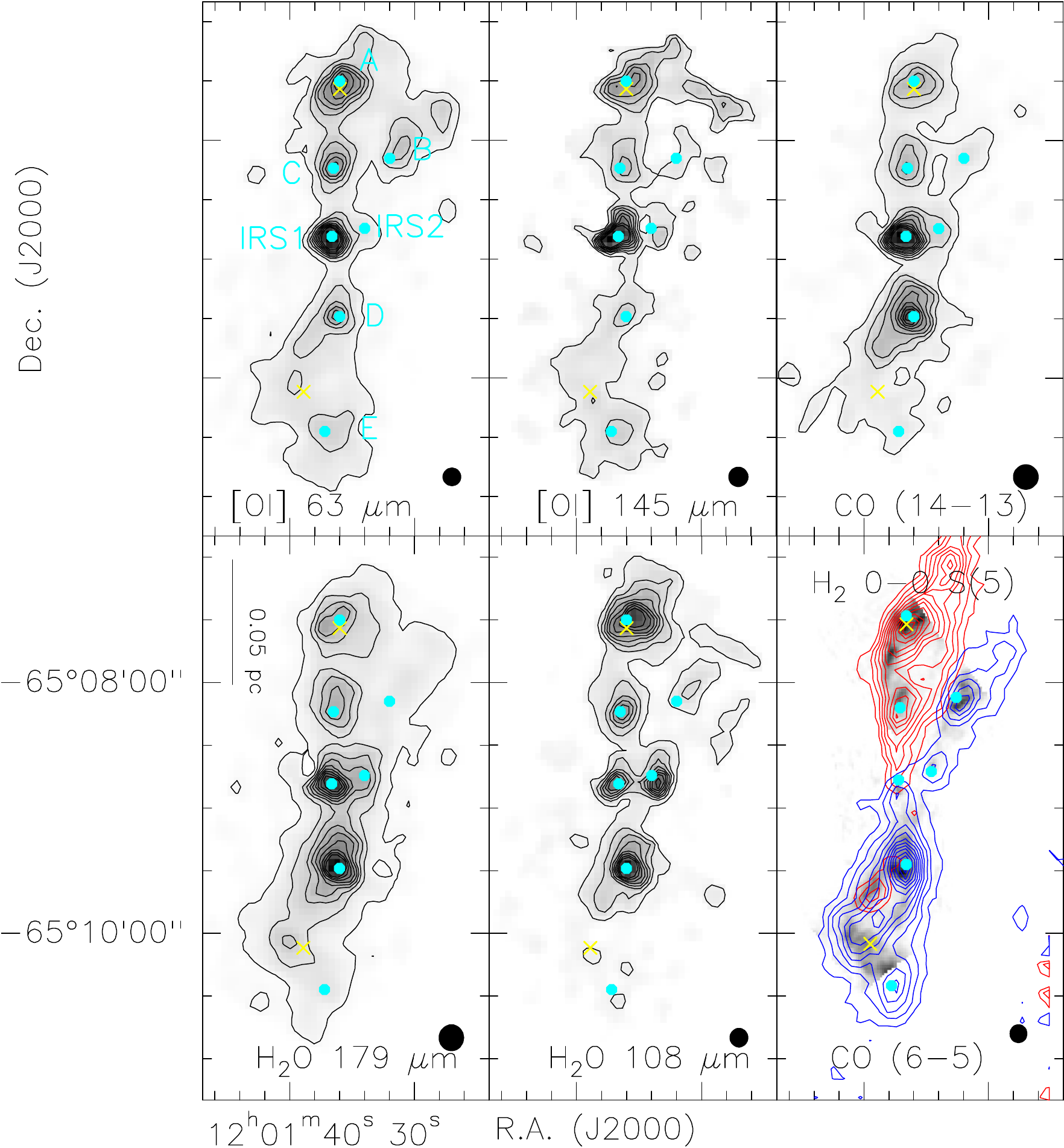}
   \caption{Integrated line maps from {\it Herschel}-PACS of the inner region of the BHR71 outflow system. The bottom right panel shows the H$_2$ 0-0 S(5) \citep{neufeld09} and the red (from 0 to 20 \kms) and blue (from $-$20 to $-$9 \kms) wings of the \cosei\, emission \citep{gusdorf15}. First level is at 3$\sigma$, namely: 2.1$\times$10$^{-5}$ \unitmap\, for \oi\, 63 \um, 3$\times$10$^{-6}$ \unitmap\, for \oi\, 145 \um, 5$\times$10$^{-6}$ \unitmap\, for \coquattordoci, 7.5$\times$10$^{-6}$ \unitmap\, for H$_2$O 179 \um, 4.5$\times$10$^{-6}$ \unitmap\, for H$_2$O 108 \um\, and 6 K \kms\, for \cosei.   
   Steps corresponds to 8$\%$ of the maximum of each line up to the 70$\%$ of the maximum. Cyan circles indicate the position of the peaks of the {\it Herschel} line emission, labelled from A to E in order of decreasing declination plus the protostellar sources IRS1 and IRS2. The two yellow crosses indicate the two positions observed with the additional pointed {\it Herschel} observations. The black circles in the bottom right part of each panel show the FWHM of the map.}
   \label{fig_maps_all}
   \end{figure*}

The BHR71 outflows system was mapped with PACS in CO (14--13), H$_2$O (2$_{21}$--1$_{10}$), H$_2$O (2$_{12}$--1$_{01}$) and [\ion{O}{i}] $^3P_0$--$^3P_1$ by using the line spectroscopy observing mode. The final extended map was built combining three 47\arcsec$\times$47\arcsec\, partially overlapping, Nyquist-sampled sub-maps (see Fig. \ref{fig_full_sistem}). Data were reduced with the HIPE\footnote{HIPE is a joint development by the {\it Herschel} Science Ground Segment Consortium, consisting of ESA, The NASA {\it Herschel} Science Center, and the HIFI, PACS and SPIRE consortia.} v9.0 package, where they were flat-fielded and flux-calibrated by comparison with observations of Neptune. The flux calibration uncertainty amounts to $\sim$ 20$\%$, while the {\it Herschel} pointing accuracy is $\sim$2\arcsec. Post-pipeline reduction steps were performed to locally fit and remove the continuum emission and to construct integrated line maps (see Fig~\ref{fig_maps_all}).

\subsection{{\it Herschel} single pointing}

Two positions (shown in Fig.~\ref{fig_full_sistem}) were observed in additional spectral lines: CO (5--4), $^{13}$CO (5--4), CO (9--8), CO (14--13) with HIFI and CO (14--13), (18--17), (20--19), (22--21), OH at 119 \um\, and [\ion{O}{i}] at 145 \um\, with PACS. The two positions correspond to local H$_2$ peaks, namely the SiO knot \citep{gusdorf11} in the northern lobe (R.A.(J2000) $12^{\rm h} 01^{\rm m} 36.00^{\rm s}$, Dec.(J2000) -65\degr07\arcmin34\arcsec) and knot 8 \citep{giannini04} in the southern lobe (R.A.(J2000)  $12^{\rm h} 01^{\rm m} 38.90^{\rm s}$, Dec.(J2000) -65\degr 10\arcmin 07\arcsec).
These observations were performed with a single pointing therefore for each line the HIFI data consist of a single spectrum while the PACS data is composed by 25 spectra combined in a 47\arcsec$\times$47\arcsec\, map, not Nyquist-sampled with respect to the point spread function (PSF). PACS data were reduced in the same way as the extended maps exposed in the previous subsection by using the chopping-nodding pipeline v10.1.0 within HIPE. HIFI data, both the Wide Band (resolution 1.1 MHz) and High Resolution (resolution 0.25 MHz) Spectrograph backends (WBS and HRS, respectively), were reduced with the standard product generation (SPG) v11.1.0 within HIPE. After consistency check-ups, the horizontal and vertical polarisation spectra are averaged in order to increase the signal-to-noise ratio. The observed lines are well resolved already at the WBS resolution therefore in our scientific analysis we will use those spectra. Calibration of the raw data onto the T$_{\rm A}^*$ scale was performed by the in-orbit system, while the spectra were converted to a T$_{\rm mb}$ scale by adopting a forward efficiency of 0.96 and the main beam efficiencies given in \citet{roelfsema12}.

In the northern pointing, the \coquattordoci\, line has been detected by both PACS and HIFI. There is a difference between the two measurements of $\sim$ 25\% that can be considered as the estimate of the intercalibration error between the two instruments.  Therefore, in the following analysis we assume a conservative error of the absolute fluxes of 30\% that takes into account both the instrumental calibration error and the error of cross calibration between the different instruments.

\section{Results}
\label{results}

\begin{table}
\caption{Knots coordinates.}             
\label{knots}      
\centering 
\begin{tabular}{c c c}  
\hline\hline                 
 Knots label & RA (J2000) & Dec (J2000) \\
 \hline
 A & 12$^{\rm h} 01^{\rm m} 36\fs0$ & -65\degr07\arcmin30\farcs0\\
 B & 12$^{\rm h} 01^{\rm m} 32\fs0$ & -65\degr08\arcmin09\farcs0\\
 C & 12$^{\rm h} 01^{\rm m} 36\fs5$ & -65\degr08\arcmin09\farcs0\\
IRS1 & 12$^{\rm h} 01^{\rm m} 36\fs6$ & -65\degr08\arcmin48\farcs5\\
IRS2 & 12$^{\rm h} 01^{\rm m} 34\fs0$ & -65\degr08\arcmin44\farcs5\\
 D & 12$^{\rm h} 01^{\rm m} 36\fs0$ & -65\degr09\arcmin29\farcs0\\
 E & 12$^{\rm h} 01^{\rm m} 37\fs2$ & -65\degr10\arcmin27\farcs0\\
 \hline
\end{tabular}
\end{table}

PACS line maps are displayed in Fig.~\ref{fig_maps_all}. It is worth noting that the PACS lines are spectroscopically un-resolved, therefore it is not possible to separate the contribution of the red and blue outflow lobes. To this aim, in the same figure we also show the emission of the \cosei\, line observed with APEX \citep{gusdorf15}, integrated over the red and blue wings that shows the relative position of the two outflows. North of the two exciting sources there are the red lobe of the IRS1 outflow and the blue lobe of the IRS2 outflow, and the two lobes are spatially well separated. The situation of the southern lobes is more complicated, with the fainter red lob of the IRS2 outflow overlapping the more extended and brighter blue lobe of the IRS1 outflow. 

The emission of the PACS lines appears to be clumpy with several individual peaks surrounded by a low level extended emission. The strongest peak is observed towards the protostellar source IRS1 ($L_{\rm bol}$=13.5 \lsun), the exciting source of the more extended outflow. The second protostellar source IRS2 ($L_{\rm bol}$=0.5 \lsun) is weaker but still visible as a local maximum in the maps. We named the five knots located in the outflow lobes from A to E in decreasing value of declination. A first series of knots, namely A, C, D and E are well aligned along a north-south straight line crossing IRS1 and have a separation of $\sim$ 40\arcsec\, in the northern lobe and $\sim$ 60\arcsec\, in the southern lobe. They likely might indicate the direction of the driving jet emitted by IRS1. Knot B is the only knot clearly associated with the blue lobe of the IRS2 outflow. No clear, strong peak along the straight line connecting B to IRS2 is present in the southern lobe suggesting that in this lobe the emission in the PACS lines should be dominated by the blue lobe of the IRS1 outflow, as also suggested by the \cosei\, map.
The emission peaks are spatially coincident in all the observed PACS lines and are roughly located at the positions of shocked spots previously identified through H$_2$ observations \citep{neufeld09}. We measured the position of the knots from the peak of \coquattordoci\, emission, apart from knot E whose coordinates are measured from the [\ion{O}{i}] 63 \um\, map. In Table \ref{knots} the coordinates of the five outflow knots and the two protostars are listed.

\begin{figure}
   \centering
   \includegraphics[width=8cm]{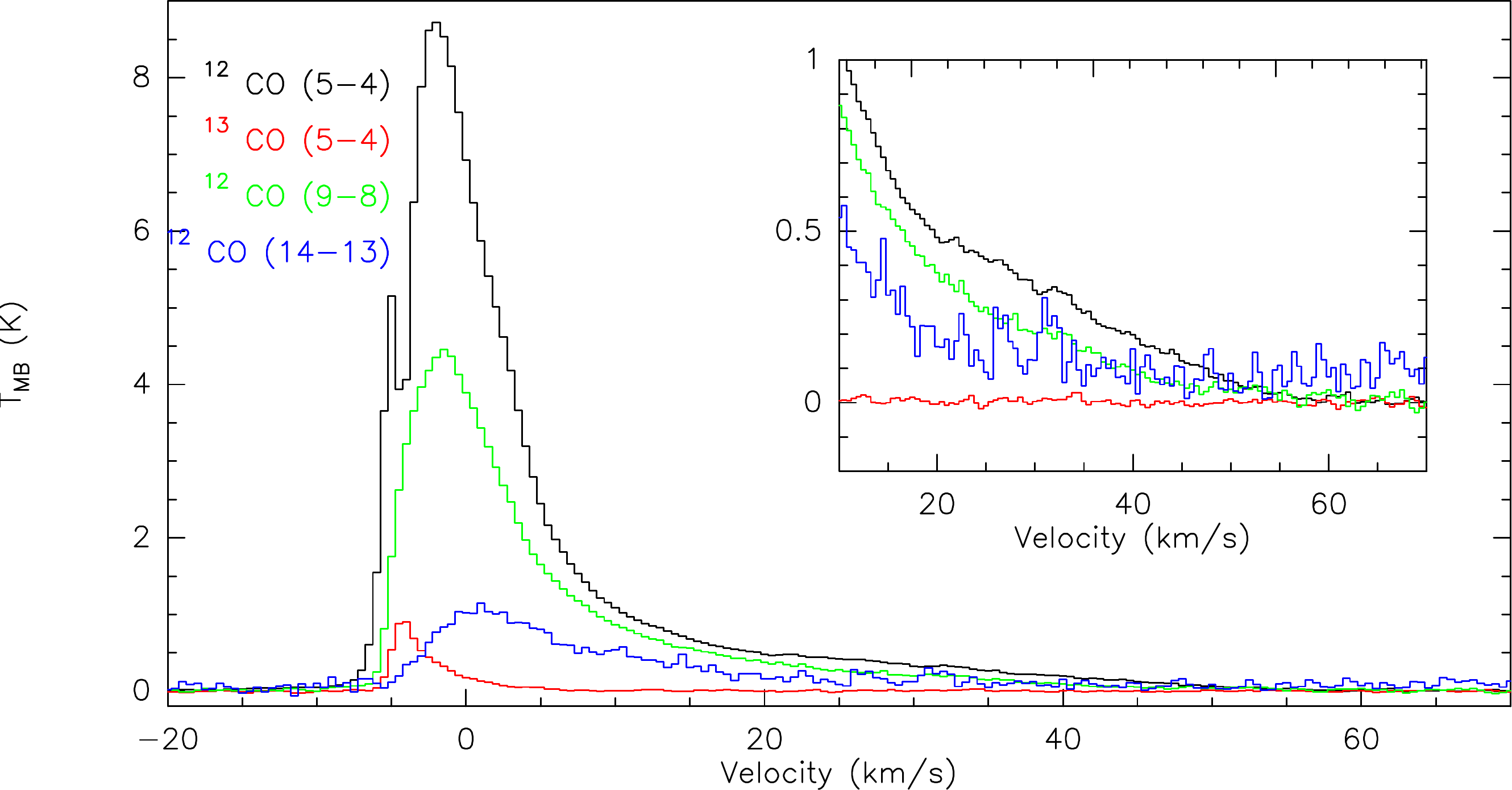}   
   \includegraphics[width=8cm]{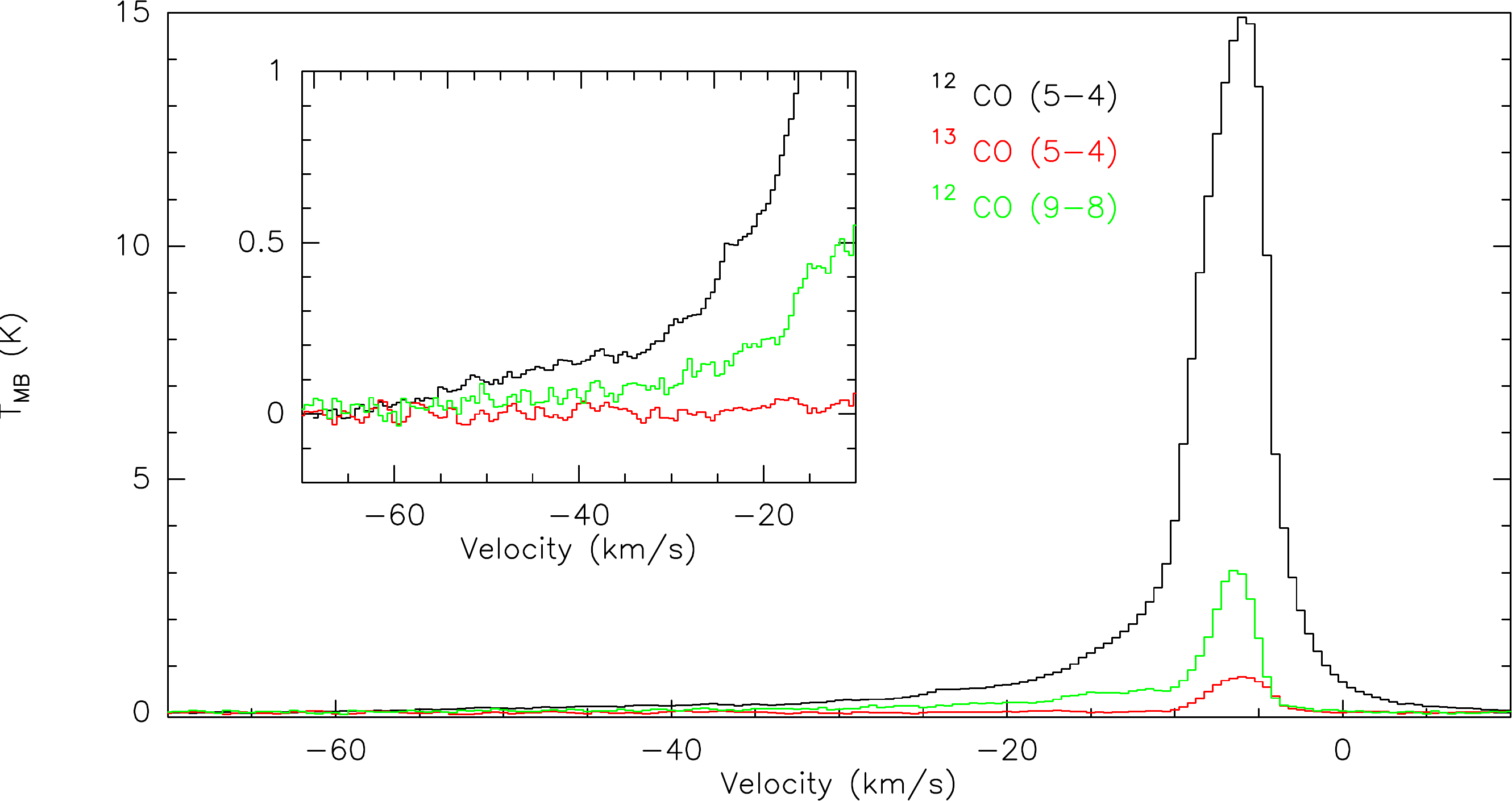}
   \caption{Plot of the lines observed with HIFI towards knot A in the northern lobe (upper panel) and about 15\arcsec\, north-east of knot E in the southern lobe (lower panel). Spectra were rebinned at a common spectral resolution of 0.5 \kms. In the insets a zoom in the lines wings is shown.}
   \label{fig_hifi_lines}
   \end{figure}

The additional single pointing observations with {\it Herschel} PACS and HIFI were acquired towards knot A in the northern lobe and about 15\arcsec\, north-east of knot E in the southern lobe, indicated by two crosses in Figs.~\ref{fig_full_sistem} and \ref{fig_maps_all}. The spectra of the four lines observed with HIFI are shown in Fig.~\ref{fig_hifi_lines}. Towards knot A, all the four lines are detected with good signal-to-noise ratio, with wings extending up to $\sim$60 \kms, a velocity higher than previously reported ($\sim$ 45 \kms, \cite{gusdorf15}). The \cocinque\, line presents a self-absorption feature at the velocity of the environmental cloud ($-$4.5 \kms, \citealt{bourke95}), observed in all CO lines up to the (6--5) \citep{gusdorf15}. From the $^{12}$CO (5--4)/\trecocinque\, ratio we measured the optical depth of the $^{12}$CO (5--4) line that results to be optically thin only in the red wing at $\varv \geqslant$ 2 \kms. The intensity peak of the CO lines moves towards higher velocity for lines with higher \jup, indicating that the low and high \jup\, lines are emitted from different layers of the shocked region. In the southern lobe pointing the lines profiles
are different, despite the blue wings extending up to a velocity ($\varv \sim -$60 \kms) similar to that observed in the red lobe. In particular, no self-absorption and no peak shifting is observed in the southern tip of the blue lobe. Moreover, the \coquattordoci\, was not detected and the line ratio \conove/\cocinque\, is lower, indicating that the excitation condition at this position is lower than in knot A. This is also confirmed by the fact that the northern pointing corresponds to a peak of the high excitation transitions observed with PACS while the southern pointing is not centred on a peak. The $^{12}$CO (5--4)/ $^{13}$CO (5--4) ratio shows that also in the southern blue lobe the $^{12}$CO (5--4) line is optically thin only in the wing, at $v< -$10 \kms. The \conove\, line in this point of the southern lobe is at velocity bluer than systemic ($\varv < -$4.5 \kms) while in the \cocinque\, line a small red wing is present. There are several possible explanations for this. The first possibility is the larger beam of the \cocinque\, line with respect to the \conove, secondly, the excitation condition of the gas in the red lobe of the IRS2 outflow could be insufficiency to populate the CO \jup=9 level, and a third possiblilty is that the emission at red velocity of the \conove\, line is under the sensitivity level of HIFI spectrograph. In all cases we can conclude that the red lobe of the IRS2 outflow does not contribute to the measured flux of the transitions with high excitation levels.

\begin{figure*}
   \centering
   \includegraphics[width=5cm]{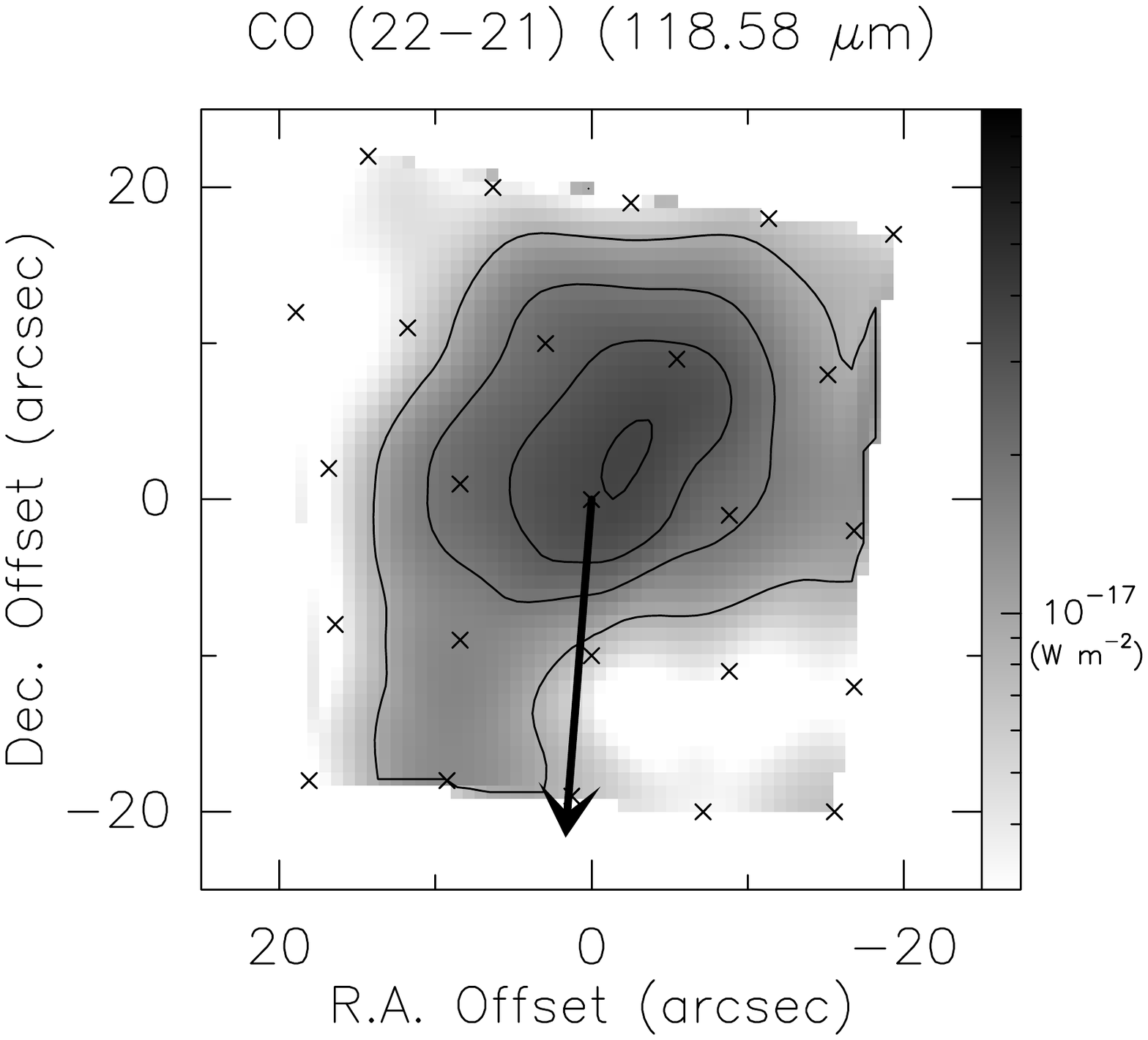}
   \includegraphics[width=5cm]{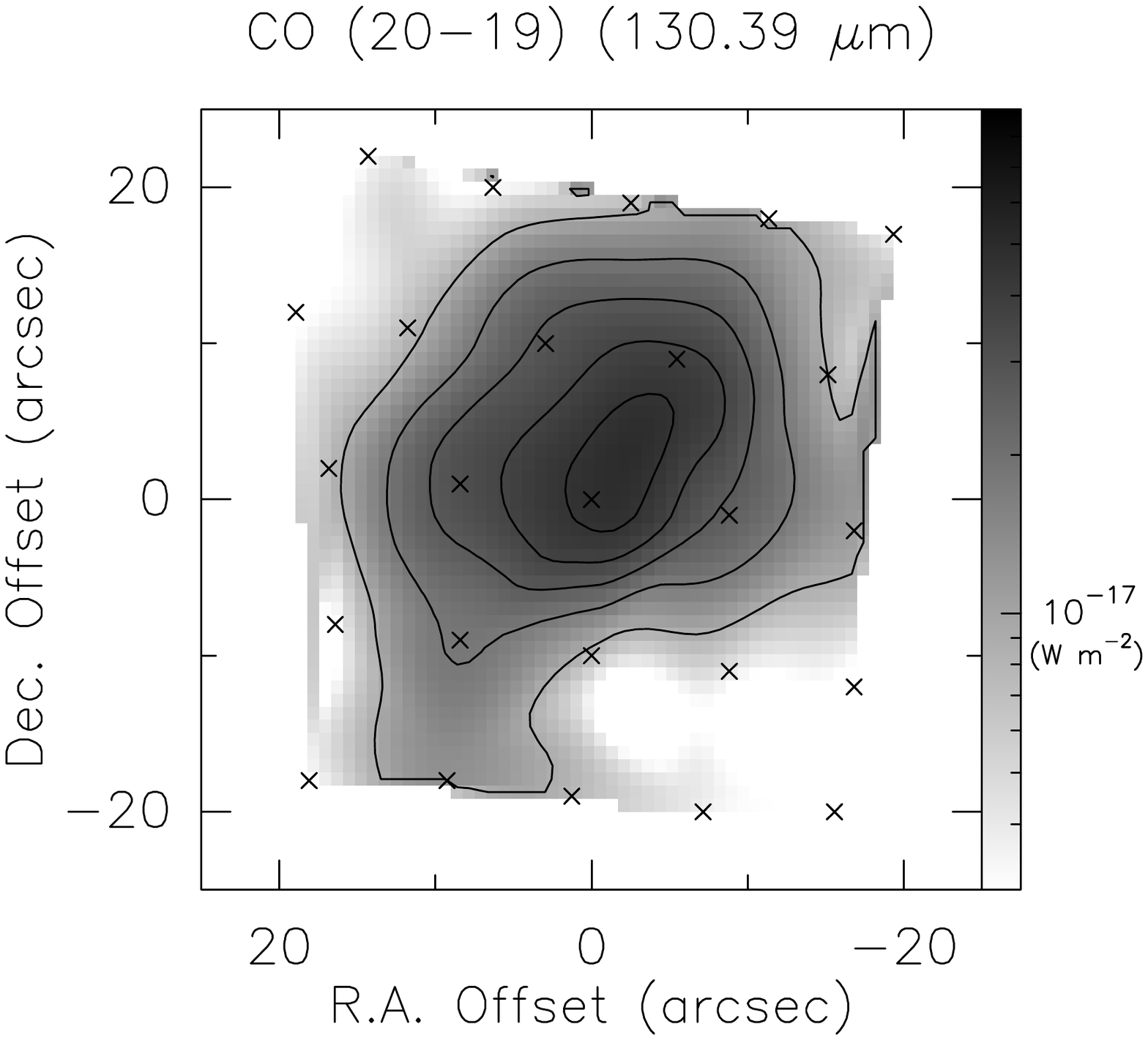}
   \includegraphics[width=5cm]{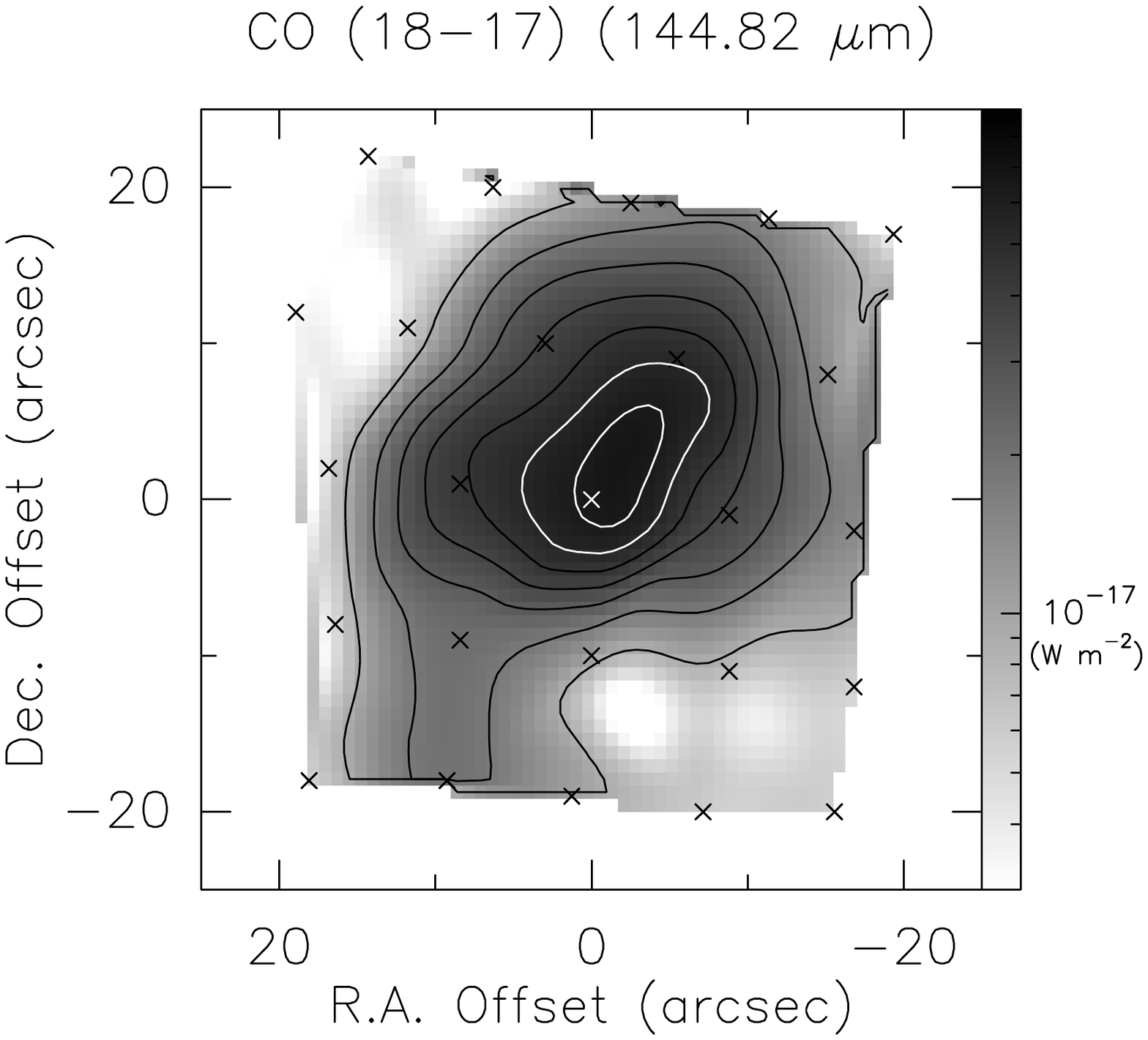}
   \includegraphics[width=5cm]{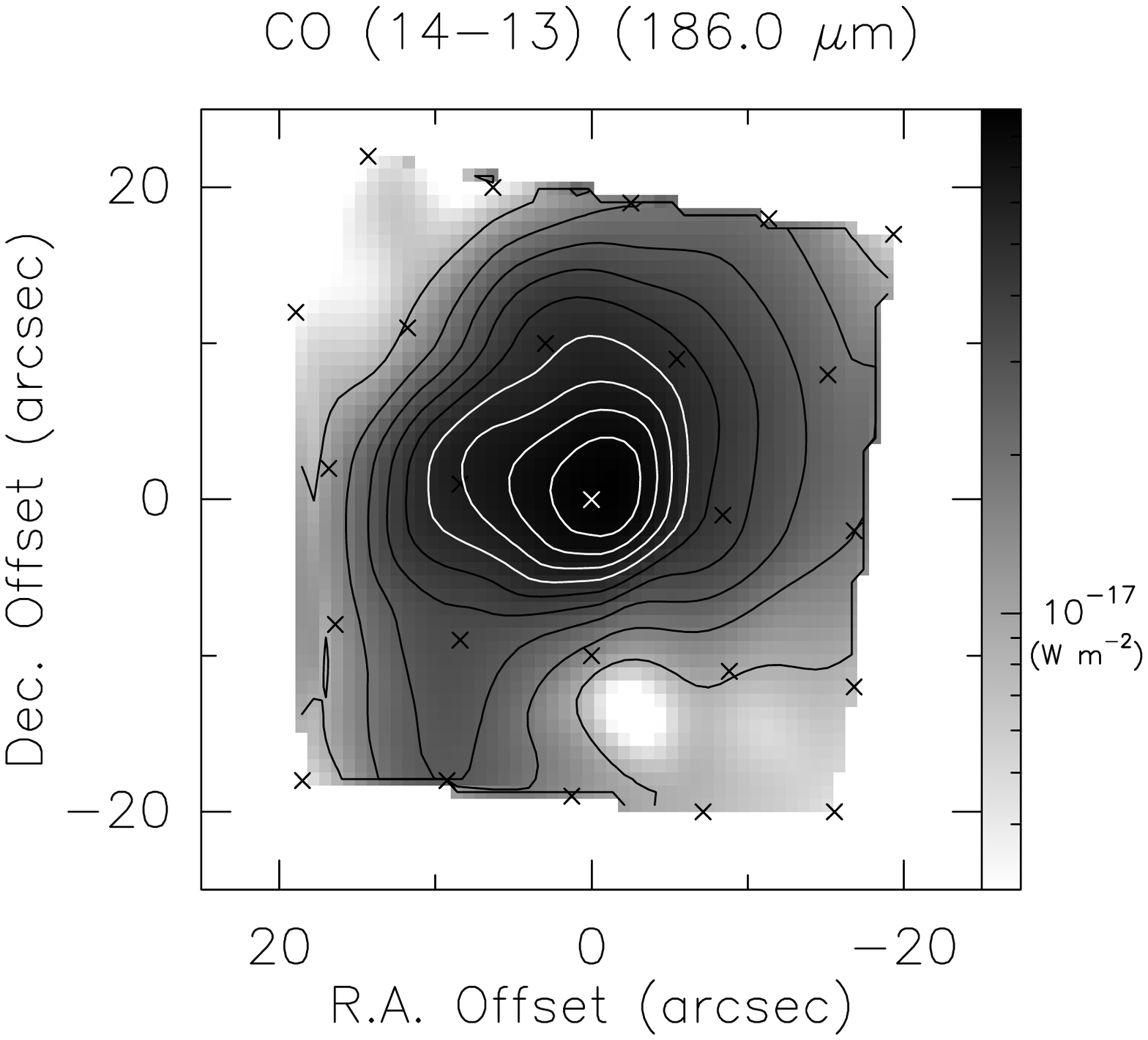}
   \includegraphics[width=5cm]{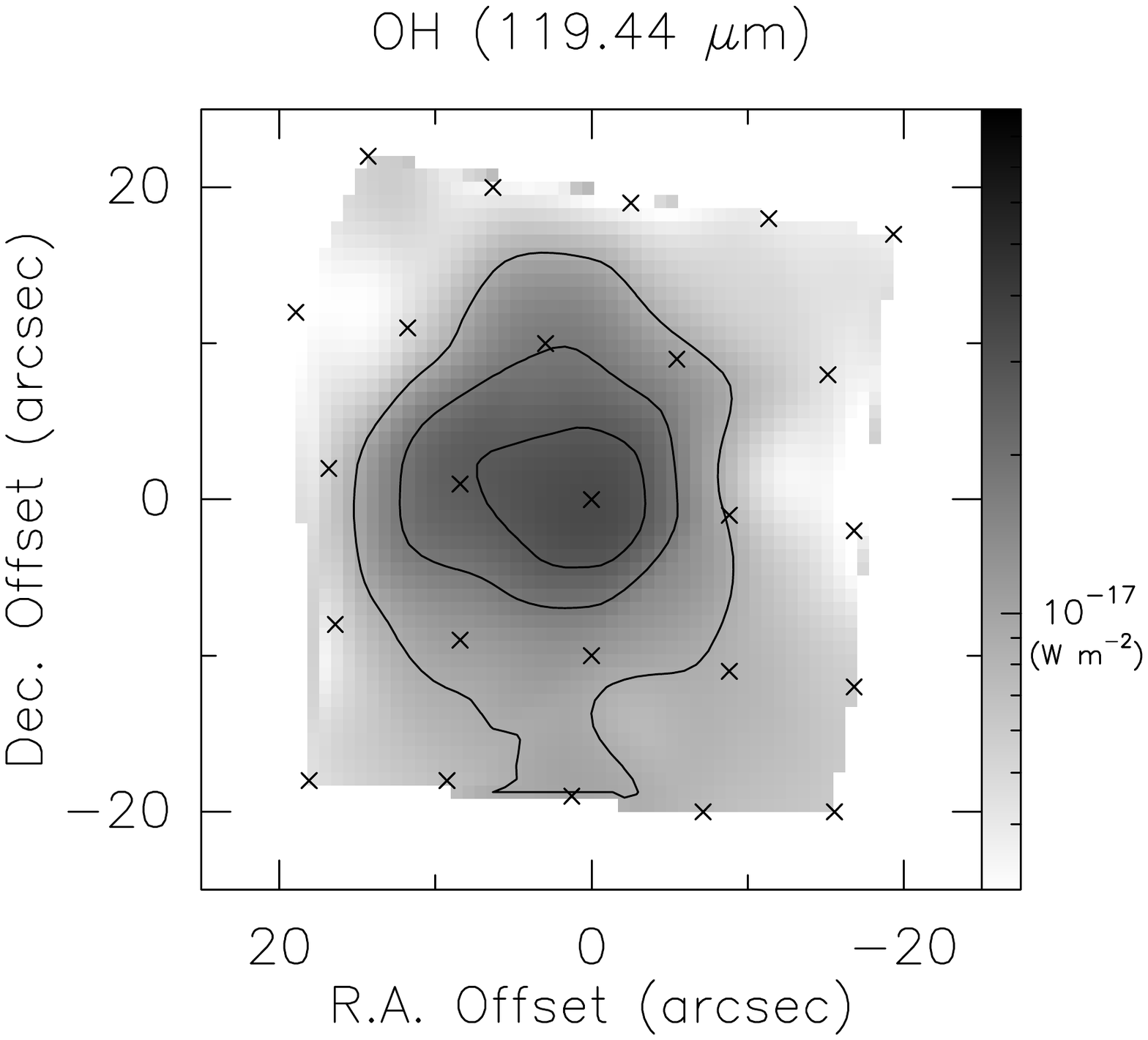} 
   \includegraphics[width=5cm]{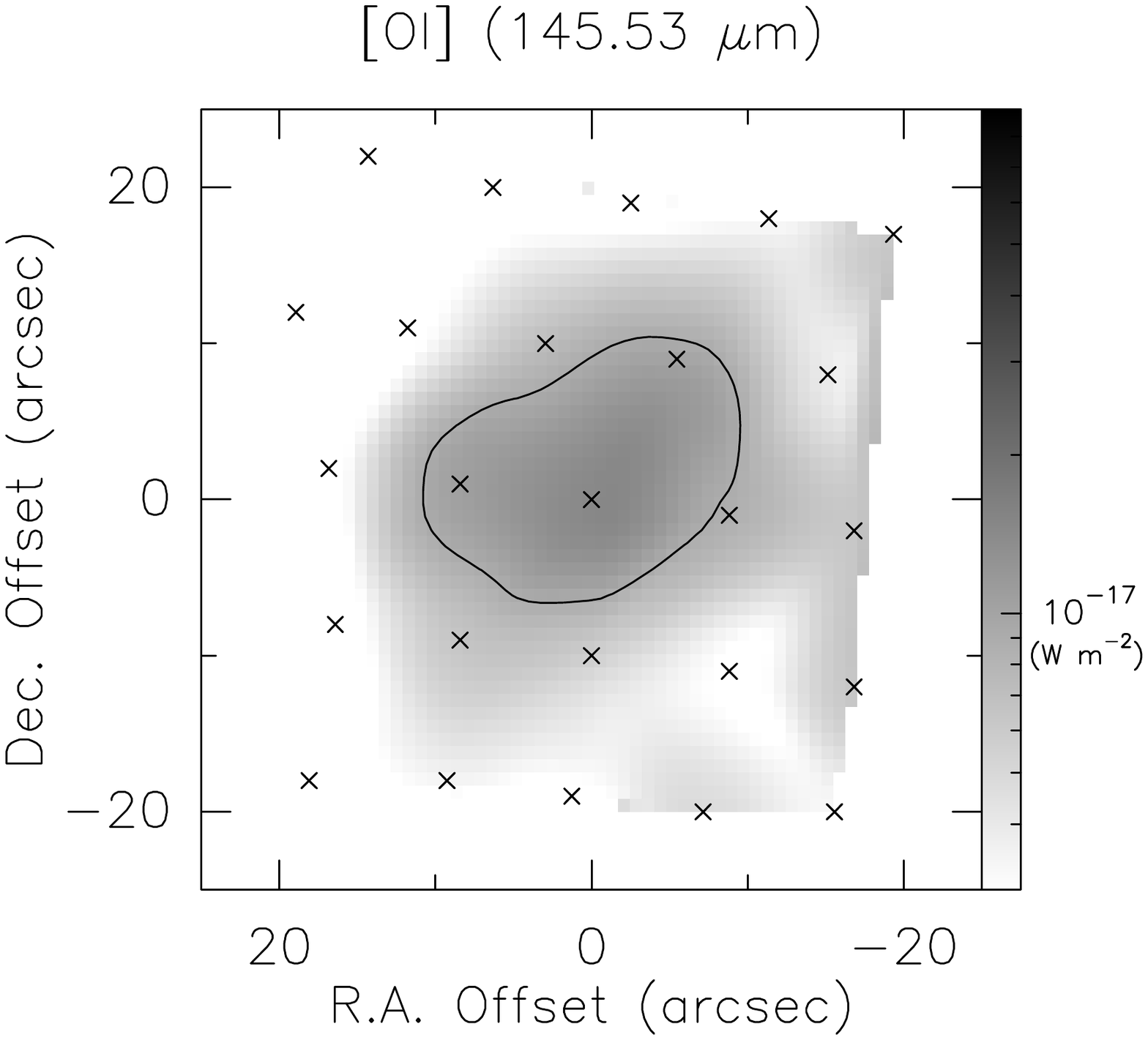}
\caption{Maps of the four CO lines, OH 119 \um\, and \oi\, 145 \um\, observed with PACS towards knot A in the red lobe of the IRS1-driven outflow. First level and level steps are 9$\times$10$^{-18}$ \wmq, corresponding to 3$\sigma$. The crosses mark the central position of the 25 spatial pixels of the PACS FOV. The arrow in the top left panel indicates the direction of the IRS1 protostar.}
\label{fig_pacs_maps_red}
   \end{figure*}

Towards the two outflow positions (indicated in Figs.~\ref{fig_full_sistem} and \ref{fig_maps_all}) we also acquired PACS maps of high \jup\, CO transitions, namely \coquattordoci, \codicotto, \coventi, \coventidue, and of the OH 119 \um\, and \oi\, 145 \um\, lines. Towards knot A, the CO lines are bright, while the OH and \oi\, are less intense (Fig.~\ref{fig_pacs_maps_red}). All the lines have a similar, partially resolved, roundish form with the peak at the centre of the map and a deconvolved circular size of 18\arcsec, indicating that they are emitted from the same extended gas component. 
Towards the southern pointing only a $\sim 3\sigma$ level \coquattordoci\, emission is detected while the other lines are not detected, confirming once again the lower excitation condition of this position.

\section{Analysis of the extended outflow}

\subsection{The distribution of the warm gas}

\begin{figure}
\centering
\includegraphics[width=8cm]{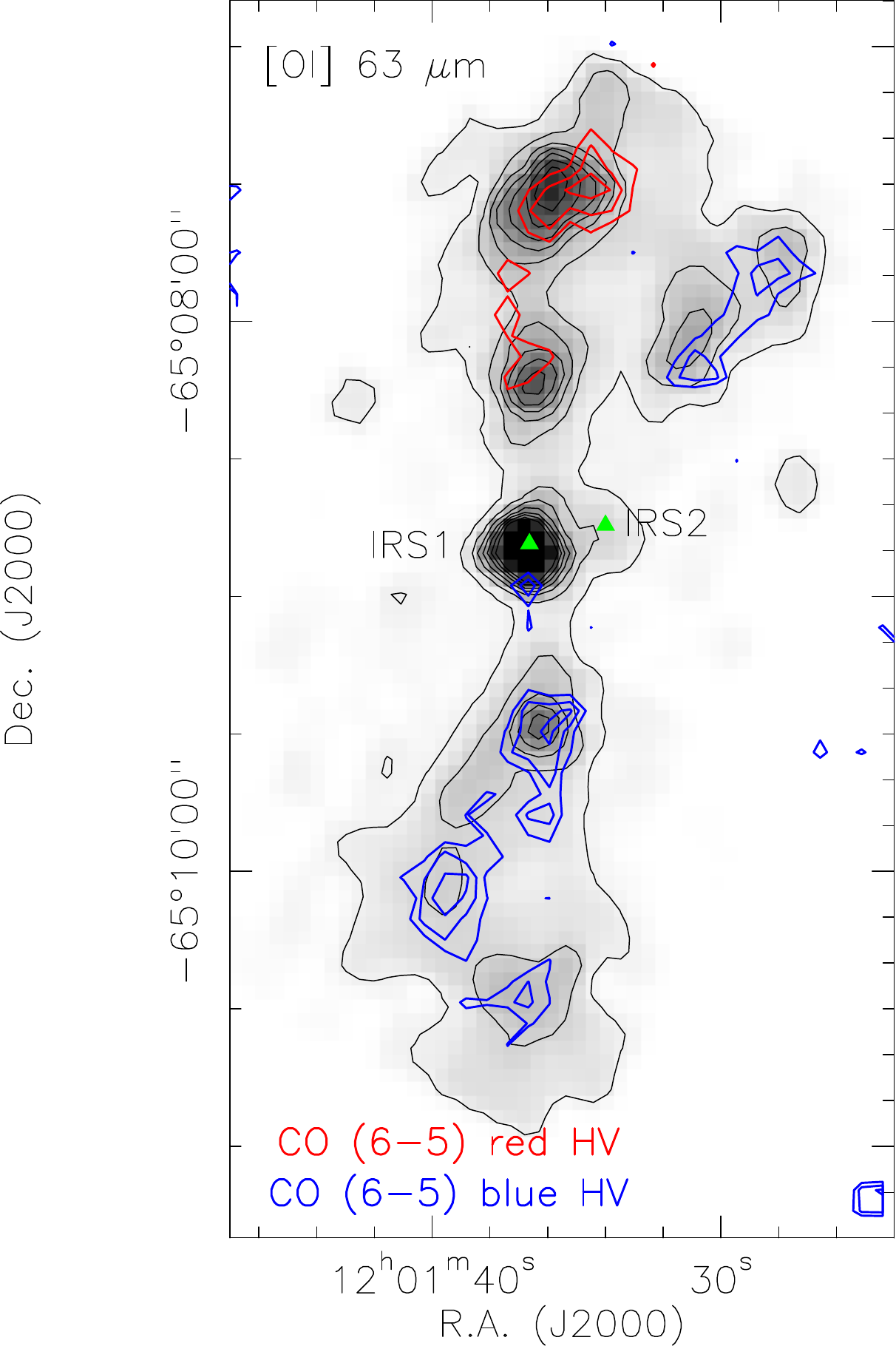}
\caption{Map of the high velocity wings of \cosei\, overlaid to \oi\, 63\um. The red wing (red line) is integrated from 10.5 \kms\, to 40 \kms\, and the blue wing (blue line) from -30 \kms\, to -19.5 \kms. CO levels are at 50\%, 70\% and 90\% of the maximum. The position of the two protostars are indicated with a triangle.}
\label{fig_hv}
\end{figure}

The far-infrared PACS lines with excitation temperatures between 80 and 580 K allow for the first time to have a complete view of the distribution of the warm gas component in the BHR71 outflow system, tracing the intermediate excitation condition between the cold gas traced by low \jup\, CO lines \citep{parise06,gusdorf15} and the hot gas traced by H$_2$ \citep{neufeld09,giannini11}.
The emission in the PACS lines shows the presence of seven bright knots surrounded by a low level emission. Two knots correspond to the two protostars that generate the outflows system and the other five knots are close to clumps of high temperature gas ($T\geqslant$ 1000 K) already observed in H$_2$ rotational transitions (see Fig.~\ref{fig_maps_all}). They are also associated with the gas moving at the highest velocity as shown by the comparison of the \oi\, 63 \um\, map with the map of the reddest (from 10.5 \kms\, to 40 \kms) and bluest (from $-$30 \kms\, to $-$19.5 \kms) velocities of the \cosei\, line wings (Fig. \ref{fig_hv}). They most likely represent shock spots where the fast jet that drives outflows impact against the lower expanding gas.

Since the PACS lines have coarse spectral resolution, we cannot spectroscopically separate the contribution of the two outflows to the total emission. This is not a problem for the northern lobes since the two outflows are spatially well separated but it could be a problem for the southern lobes that partially overlap. However, the spectral profile of the CO lines observed with HIFI at the tip of the southern lobe close to knot E, shows that the percentage of the emission at velocity redder than the systemic velocity ($-$4.5 \kms) with respect to the total emission decreases for increasing \jup, being only the 3$\%$ for the \conove\, transition. Therefore we can expect an even lower percentage for the \coquattordoci\, line and for the other PACS lines from [\ion{O}{i}] and water, that trace the same warm gas component, as shown from Fig.~\ref{fig_maps_all} for BHR71 and as also observed in other outflows \citep{santangelo13,busquet14}.In conclusion, at the tip of the southern lobe the contribution of the red lobe of the IRS2 outflow to the PACS lines is negligible. The situation in the southern positions closer to the two protostars could be different since in the uppest \jup\, available CO line, the \cosei\, observed with APEX, the emission in the red wing is still significant. However, towards the bright knot D the similar morphology of the blue wing of the \cosei\, and the PACS lines (Fig.~\ref{fig_maps_all}) shows that even in this knot the contribution of the red lobe of the IRS2 outflow to the PACS lines should be negligible.

\begin{figure}
   \centering
   \includegraphics[width=8cm]{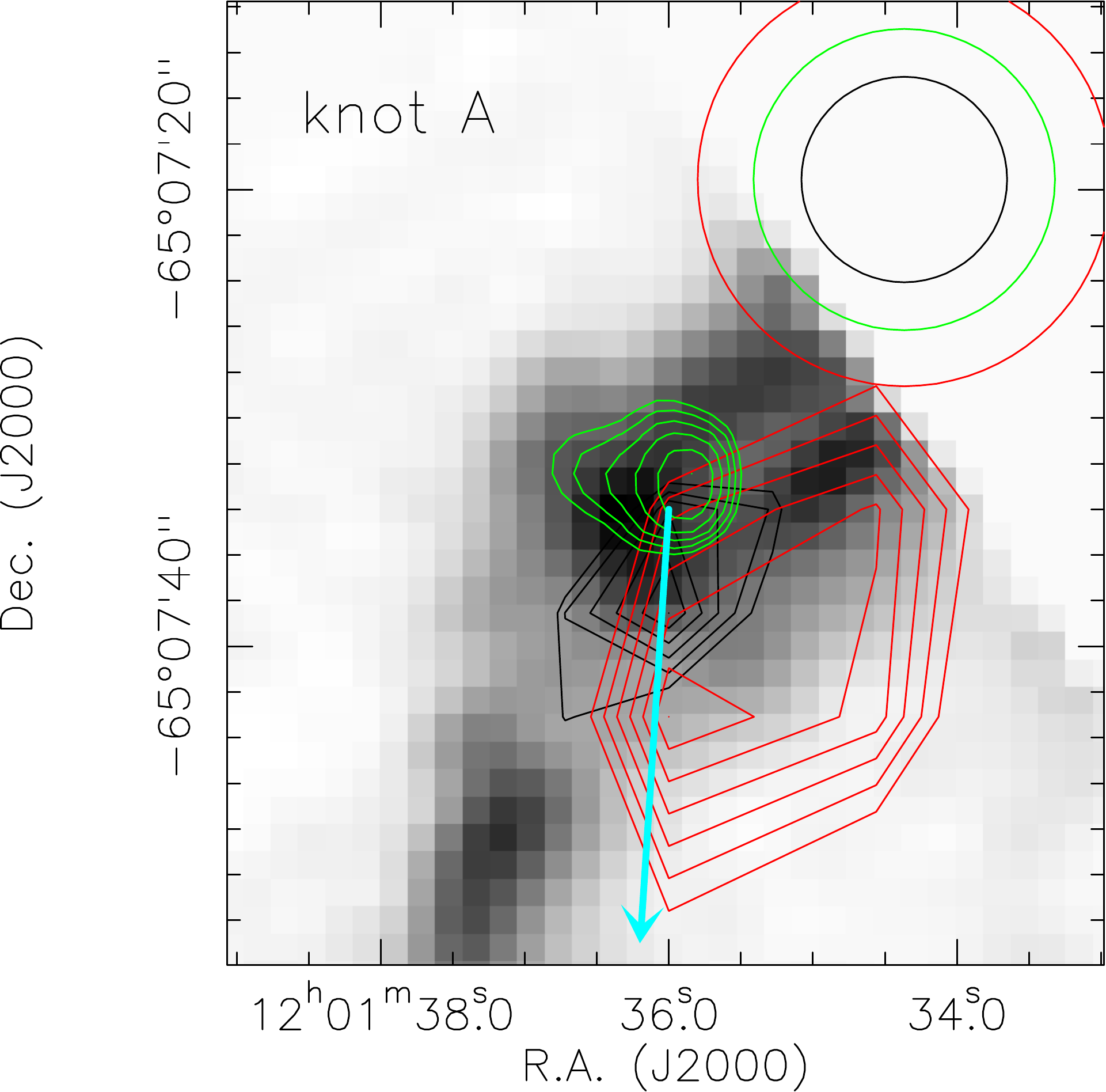}
   \includegraphics[width=8cm]{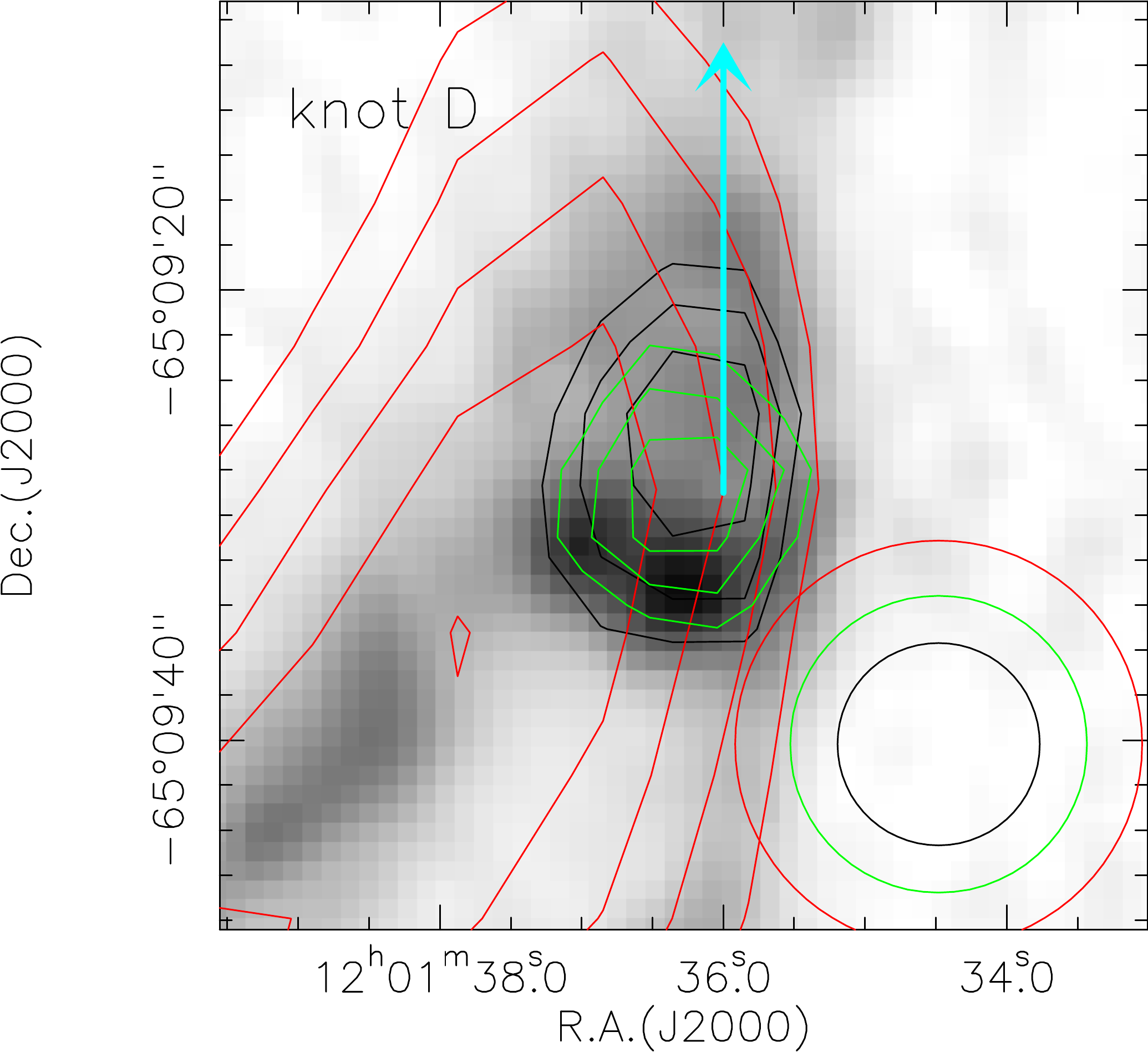}   
   \caption{Map of the H$_2$ 0-0 S(5) line (grey-scale map) compared with CO (3--2) (red line), CO (6--5) (black line) \coquattordoci\, (green line). Upper panel is centred on knot A, lower panel on knot D. Contours start at 90\% of the peak value in order to highlight the position of the peak of the emission. The circles show the HPBW of the different lines. The arrows indicate the direction of IRS1.}
   \label{fig_co_vari}
   \end{figure}

In Sect.~\ref{results} we already noticed that towards knots A the peak velocity of the CO lines moves to higher velocity for lines with higher \jup.
In Fig.~\ref{fig_co_vari} we compare the spatial morphology of \cotre, \cosei\, and \coquattordoci\, emission in knot A  and we find that the peak of the CO emission moves towards north for CO lines with higher \jup, with a separation of about 20\arcsec\, between the \cotre\, and the \coquattordoci\, peaks. Moreover, the peak of the highest excitation CO line is the closer to the peak of H$_2$ 0-0 S(5) ($E_{\mathrm{up}}$ = 4586 K) emission tracing hot gas. The peak shifting is along the straight-line connecting knots A, C and D and crossing IRS1, suggesting that the shift is linked to the bow shock generated by the impact of the high-velocity jet against the slower moving ambient shell.
The situation in the bright knot D of the southern lobe is different (see Fig.~\ref{fig_co_vari}), with the low \jup\, CO (3--2) line having the classical morphology of the cavity walls and the higher \jup\, CO (6--5) and (14--13) showing a roundish structure with spatially coincident peaks. 

A displacement in the emission peak of the CO lines with increasing \jup\, was also observed towards the flow containing the Herbig-Haro object HH54 \citep{bjerkeli14}. In HH54, however, the gradient goes in the direction of the exciting source, that is, in the opposite direction with respect what we observe in knot A of BHR71. \citet{bjerkeli14} interpreted the observed displacement of the CO lines peak as the consequence of the presence of an under-dense, hot clump upstream of the HH54 shock wave, traced by a well defined distinct gas component in the spectra of the low \jup\, CO lines. Although no distinct velocity components are present in the CO spectra of the BHR 71 outflow, the observed displacement of the location of the emission maximum of CO with increasing \jup\, in our source could be attributed to different layers of shocked gas impacting an ambient medium with a significant stratification of particle density. The difference in the morphology of the CO lines emission between knots A and D is probably due to the different environmental conditions in which the two shocks have been originated and/or to geometrical effects that could change the angle of view under which the two shocks are observed. In fact, as clearly visible in Fig. \ref{fig_full_sistem} that compares the outflow traced by \cotre\, with an optical image of the region at 0.54~\um, the outflow is breaking out from its parental Bok globule with more than half of the blue southern lobe projected outside the obscured region while in the red northern lobe only the very final part is projected outside the globule. Although it is more difficult to understand what is really happening on the rear side of the globule, the measured much larger mass and kinetic energy of the red lobe with respect the blue lobe \citep{bourke97} confirms what Fig. \ref{fig_full_sistem} suggests.
Our {\it Herschel}-PACS map covers the part of the northern lobe still fully embedded in the obscured and denser medium inside the molecular dark cloud while the final part of the mapped portion of the southern lobe is propagating outside the cloud.
In this respect knots A and D are in a different environment: the northern knot A is located completely inside the cloud while the southern knot D is located at the border of cloud, likely in a less dense medium. Moreover, while knot A is associated to only the red lobe of the IRS1 outflow in the position of knot D two different lobes coexist.  

\begin{figure}
   \centering
   \includegraphics[width=8cm]{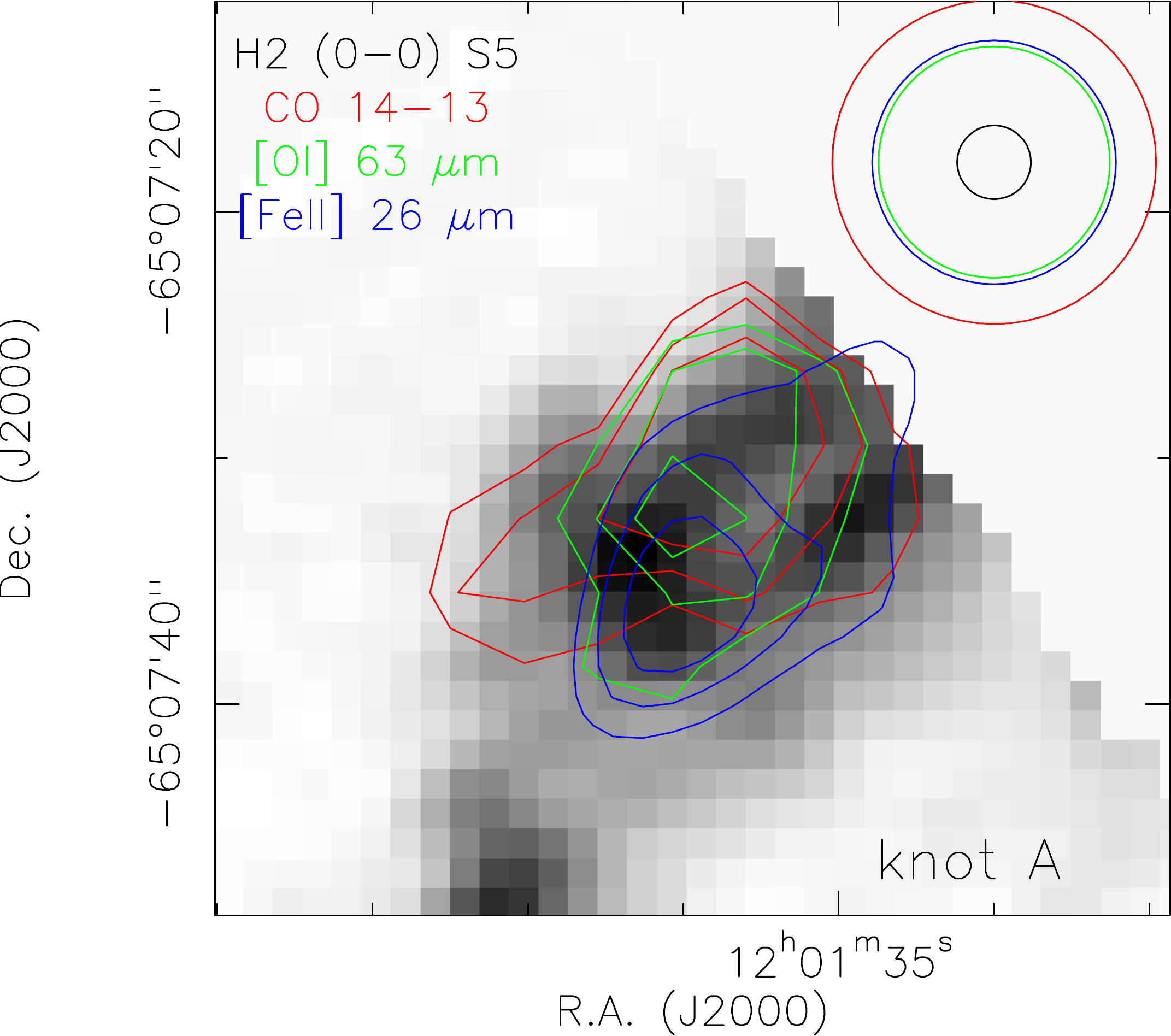}
   \caption{Map of the H$_2$ 0-0 S(5) line compared with \coquattordoci\, (red line), \oi\, 63 \um\, (green line), $[\ion{Fe}{ii}]$ 26 \um\, (blue line) towards knot A. Contour levels is 70\%, 80\% and 90\% of the local maxi ma of each line. The circles show the HPBW of the different lines.}\label{fig_zoom_red}
   \end{figure}

Remarkably, towards knot A the peak of the high \jup\, CO corresponds to the peak of atomic lines such as \oi\, 63 \um\, and $[\ion{Fe}{ii}]$ 26 \um\, and of the H$_2$ rotational lines (Fig.~\ref{fig_zoom_red}) as also observed in other bright, outflow shocks such as L1157-B1 and HH54 \citep{benedettini12, bjerkeli14}. This implies that the presence of a dissociative and ionising shock but also that the jet material is already partly molecular, or molecules reformation is efficient after the passage of the shock front. A model of the shock towards knot A and D is presented in Sect.~\ref{sect_shock_mod}.

\subsection{Line ratios}
\label{sect_line_ratio}

\begin{figure*}
\includegraphics[width=6.5cm,angle=90]{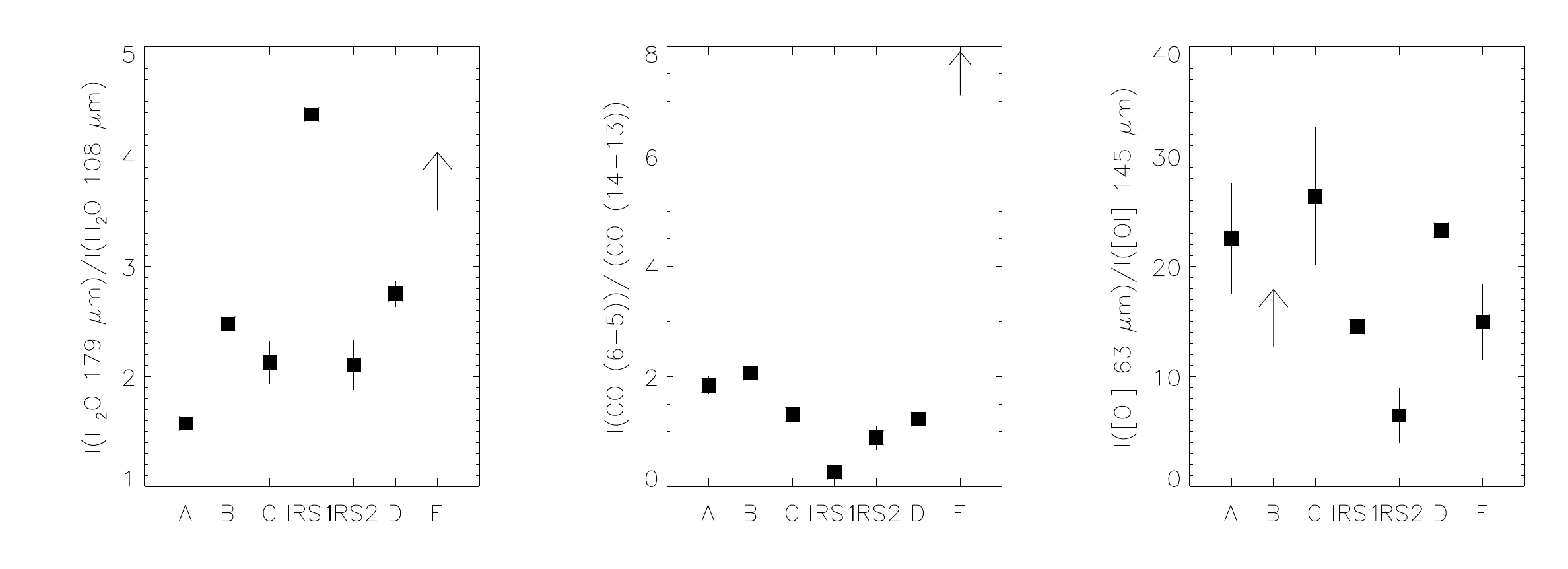}
\caption{PACS lines ratios of H$_2$O 179\um/108\um\, (left panel), CO (6--5)/(14--13) (middle panel) and \oi\, 63\um/145\um\, (right panel) in the seven infrared knots. Arrows represent lower limits.}
\label{fig_line_ratio}
\end{figure*}

With only two lines per species it is impossible to constrain the physical parameters of the emitting gas. However, we can derive some general indications on if and how the physical and excitation condition of the gas change along the outflow, comparing the ratios of lines of the same species in the seven infrared knots. In order to avoid different beam filling effects we smoothed all the PACS maps at the common spatial resolution of 13\farcs2 and extracted the brightness in the position of the seven {\it Herschel} knots. In Fig. \ref{fig_line_ratio} we show the H$_2$O 179\um/108\um, the CO (6--5)/(14--13) and the \oi\, 63\um/145\um\, ratios in the seven knots along the outflow.
The ratio between the two water lines is quite constant in all the knots but IRS1 where it is higher of about a factor of two. A higher H$_2$O 179\um/108\um\, is an indication of a lower particle density or/and lower water column density. IRS1 is also the knot with the lower CO (6--5)/(14--13) ratio which indicates a much higher temperature ($T>$ 1200 K) for this source. This is not unexpected since towards very young protostars a hot CO component was usually observed, traced by  CO lines with excitation temperature higher than 1500 K (e.g. \citealt{dionatos13}; \citealt{vankempen10}; \citealt{karska13}). In particular, towards IRS1 \citet{ karska13} detected the CO (24--23) line, indicating high temperature for the gas surrounding the protostar. The CO (6--5)/(14--13) ratio shows a decreasing trend from the northern lobe knots to the southern lobe knots and this is indicative of a lower temperature at the southern tip of the outflow that confirms our results on the lower excitation condition presented in the previous sections. 

The ratio of the two \oi\, lines gives an estimate of the hydrogen particle density. We compared our observed ratios with theoretical line intensity ratios calculated from the non-LTE code also used in \citet{nisini15} that considers the first five levels of oxygen. The three main peaks of the IRS1 outflow have ratio higher than 22, indicative of density between 10$^4$ -- 10$^5$ \cmtre\, while on sources and at the southern peak of the outflow (knot E) we measured a slightly lower ratio (15), indicating that in the latter positions the density could be lower. It is worth noting that the lowest ratio of 6.5 observed towards IRS2 is not compatible with model predictions. Unpredicted low \oi\, 63\um/145\um\, ratios have been already observed in several sources \citep{liseau06} and interpreted as due to self-absorption of the 63 \um\, line. This interpretation was corroborated by the recent observation of a spectroscopically resolved \oi\, 63 \um\, line towards a high-mass source which shows a very prominent absorption \citep{leurini15}. The self-absorption of the \oi\, 63 \um\, line could be the reason of the extremely low 63\um/145\um\, ratio observed towards IRS2. Indeed IRS2 has been classified as a Class 0 protostar and it is likely at an earlier evolutionary stage with respect to IRS1 \citep{chen08} therefore it should be deeply embedded in its dusty envelope whose extinction capability is still efficient at 63 \um\, but it does not affect lines at higher wavelengths. We cannot exclude the possibility  that the \oi\, 63 \um\, line towards IRS1 is also partially absorbed therefore the density inferred form the \oi\, lines ratio should be considered a lower limit.

\section{Modelling of the CO lines in knot A}

For the brightest point of the red lobe of the IRS1 outflow (knot A) we have measurements of CO lines from \jup = 3 to \jup = 22 that allows us to make a deep analysis of the physical conditions present in this pure shock position. The region has already been modelled in several papers \citep{gusdorf11,gusdorf15}, here we add the new contribution of the high \jup\, CO lines observed by PACS that trace a gas component not traced by the transitions modelled by previous works.

\subsection{Rotational diagram}

\begin{figure}
   \centering
   \includegraphics[width=7cm,angle=90]{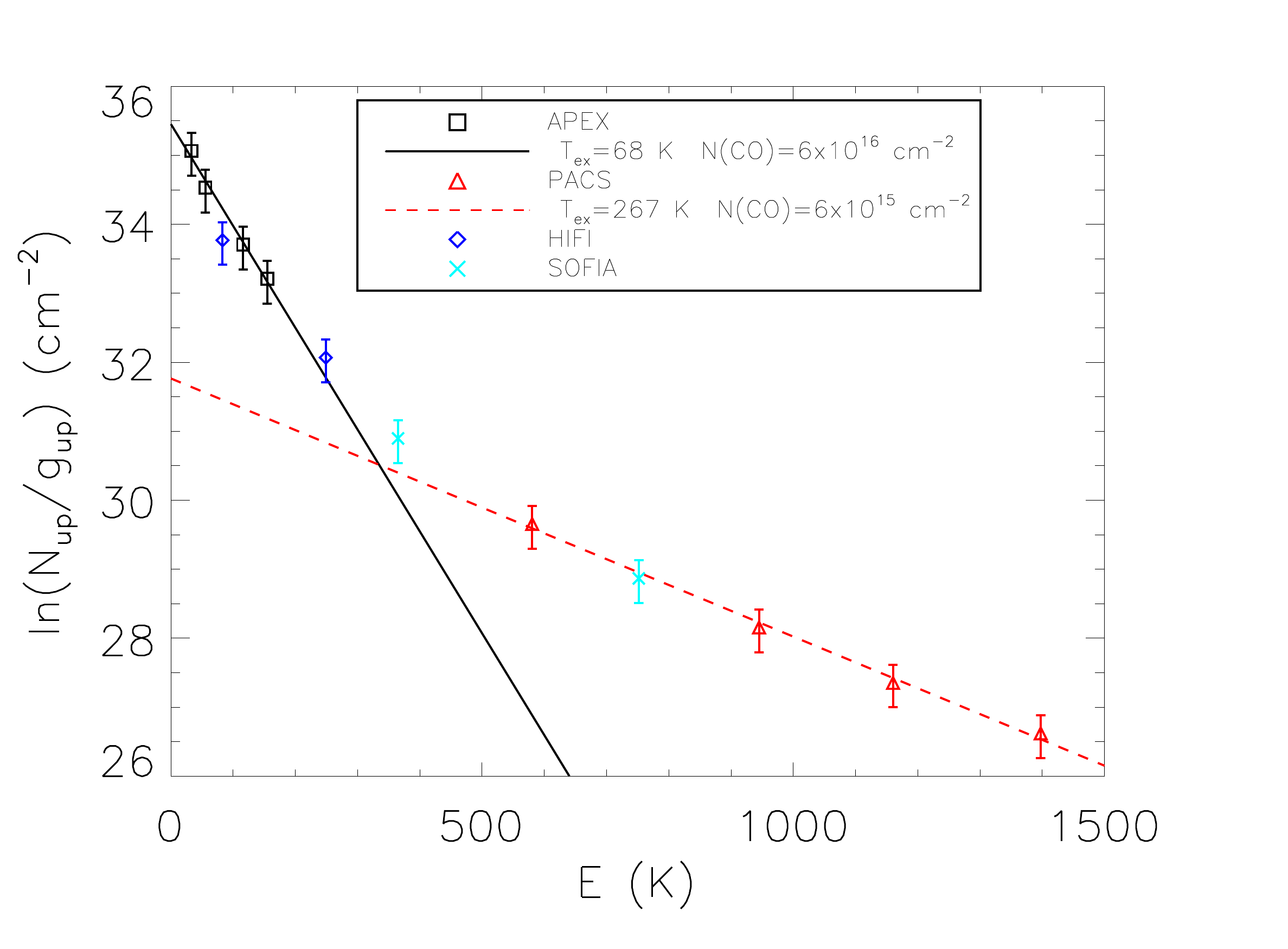}
   \caption{Rotational diagram of CO lines. For the low \jup\, lines the derived rotational temperature and column density are values averaged in the 24\arcsec\, beam, assuming a filling factor 1. For the high \jup\, lines a size of 18\arcsec\, is assumed.}
   \label{fig_corot24}
   \end{figure}

We calculated the rotational temperature from the rotational diagram of the CO lines, which gives a lower limit to the kinetic temperature if the gas is not in local thermodynamic equilibrium (LTE).
In order to enlarge our data set as much as possible we also used the low \jup\, CO line observed with APEX and SOFIA \citep{gusdorf15}. We smoothed all the CO maps at the common resolution of 24\arcsec. Four CO lines, namely \cocinque, \conove, \coundici\, and \cosedici, however, have been observed only with a single pointing and with a different beam (except the \coundici\, that was observed with a beam of 24\arcsec) and are not used in the fitting. In building the rotational diagram we took into consideration the measured size of the CO emission, assuming that lines with \jup\, from 3 to 11 are extended over the beam of 24\arcsec while the lines with \jup\, from 14 to 22 have a size of 18\arcsec\, (see Sect. \ref{results}). For the spectrally resolved lines observed with APEX and SOFIA the line flux was calculated integrating the emission in the velocity range between -4.5 \kms\, and 50 \kms\, as in \citet{gusdorf15}, while for the unresolved {\it Herschel}-PACS lines we used a Gaussian fit to the line profile.
We attribute to each line flux an uncertainty of 30\% in order to take into consideration the intercalibration error between different instruments. The rotational diagram is shown in Fig.  \ref{fig_corot24}.

Two components are clearly present: a low excitation component fitting the low \jup\, lines and a higher excitation component fitting the high \jup\, lines. To derive the column density and the rotational temperature of the two components we fitted two sets of lines separately, grouping the lines measured with the same instrument, i.e. the APEX (\jup=3, 4, 6, 7) and the PACS lines (\jup=14, 18, 20, 22), respectively. Indeed the point at which the slope of the fitting stray-line changes is between the two sets of transitions. The best fit parameters are: $T_{\mathrm{rot}}$=68 K and $N$(CO)=6$\times$10$^{16}$\cmdue\, for the low \jup\, APEX lines and $T_{\mathrm{rot}}$=267 K and $N$(CO)=6$\times$10$^{15}$\cmdue\, for the high \jup\, PACS lines. Note that the HIFI and the SOFIA measurements, not used in the fitting itself, are in agreement with the fitting results.
It is worth noting that the low \jup\, lines, at least the ones with \jup$\leq$5, are not optically thin at all velocities so that column density derived from the rotational diagram represents a lower limit and the rotational temperature an upper limit. The two rotational temperatures are similar to those found in other shock positions along protostellar outflows as L1157 and Cep E \citep{benedettini12,lefloch12,lefloch15}.

\begin{figure*}
   \centering
   \includegraphics{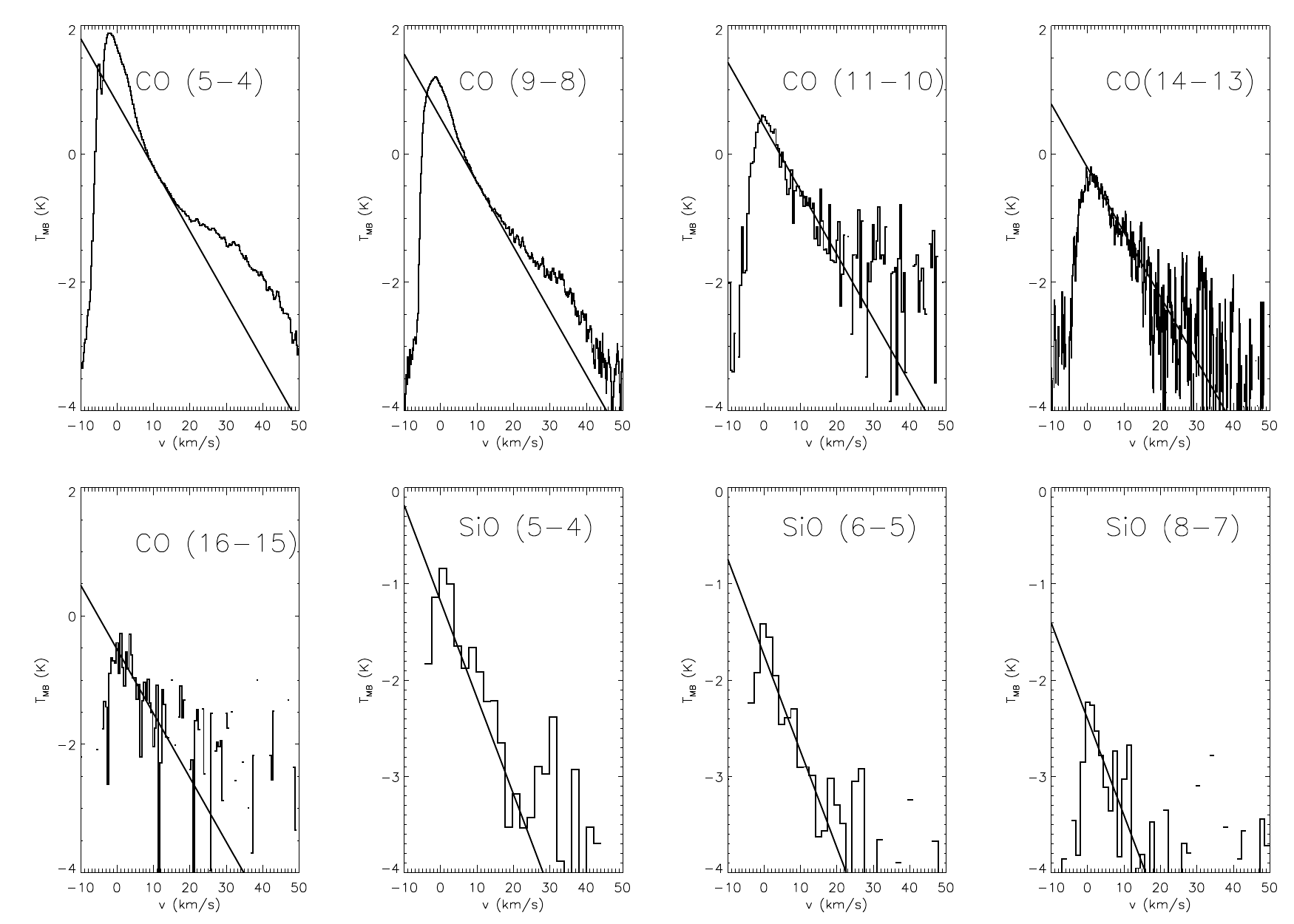}
   \caption{Exponential fit of the line profile for CO and SiO lines towards knot A. The line represent the fit to the line profile with the exponential low $I(\varv) = I$(0) exp(-$\varv$/10 \kms).}
   \label{fig_fit_profile}
   \end{figure*}

These two different gas components show a different line profile. In fact, the spectral profile of the \coquattordoci\, and \cosedici\, lines is well described by a single exponential law $I_{CO}(\varv) = I_{CO}$(0) exp(-$\varv/\varv_0$) with the same slope $\varv_0$= 10 \kms\, for the two lines (Fig. \ref{fig_fit_profile}), while the lower \jup\, lines have a more complex profile indicating that different gas components contribute to the total emission. A similar slope ($\varv_0$= 12 \kms)  has been found for the profile of the high \jup\, CO lines in the B1 shock of the L1157 outflow \citep{lefloch12} and also these authors attribute the emission of the CO line with \jup$\geqslant$13 to shock excited gas. Another indication that the high \jup\, CO lines are dominated by the high excitation gas related to the shock is that also the line profile of SiO \citep{gusdorf11}, a species introduced in the gas phase by the shock through sputtering from the dust grains or vaporisation in grain-grain collisions (e.g. \citealt{guillet09}) is well described by the same exponential law (Fig.~\ref{fig_fit_profile}). 

\subsection{LVG analysis}

\begin{table*}
\caption{Parameters of the LVG fitting of the CO lines towards knot A at a resolution of 24\arcsec.  }   
\label{tab_lvg}      
\centering              
\begin{tabular}{c c c c c}   
\hline\hline            
  Component & $N$(CO)  & size & $T$  & $n$(H$_2$) \\
            & \cmdue  &  \arcsec &  K & \cmtre \\  
\hline
  cold      &2$\times$10$^{16}$ -- 6$\times$10$^{16}$ & $>$24 & $\sim$ 80 & 3$\times$10$^5$ -- 4$\times$10$^6$ \\
  warm      & 3$\times$10$^{15}$ -- 9$\times$10$^{15}$ & 18 & 1700 -- 2200 & 2$\times$10$^4$ -- 6$\times$10$^4$ \\
\hline              
\end{tabular}
\end{table*}

   \begin{figure}
   \centering
   \includegraphics[width=9cm]{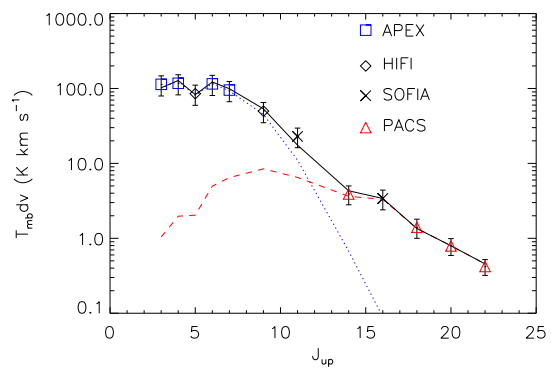}
   \caption{Best fit of the LVG model for CO lines in knot A considering two components. The parameters of the first component (dotted line) are: $N$(CO)=6$\times$10$^{16}$ \cmdue, size=24\arcsec, $T$=70 K, $n$(H$_2$)=4$\times$10$^6$ \cmtre: they are the physical parameters of the low excitation gas component averaged inside the 24\arcsec\, beam centred towards the knot A position; this component is, however, more extended than the beam. The parameters of the second component (dashed line) are: $N$(CO)=3$\times$10$^{15}$ \cmdue, size=18\arcsec, $T$=2200 K, $n$(H$_2$)=4$\times$10$^4$ \cmtre. The solid line represents the sum of the two components.}
   \label{fig_lvg_2comp}
   \end{figure}

We compared the CO lines emission with a grid of Large Velocity Gradient (LVG) models calculated with the \citet{ceccarelli03} code. The molecular data were taken from the BASECOL\footnote{http://basecol.obspm.fr} database \citep{dubernet13} and the line width (FWHM of the line profile) was set to a fixed value of 9 \kms, as measured in the HIFI spectra. We consider the first 41 levels corresponding to energy levels up to 2000 K. 
The explored range in CO column density is from 4$\times$10$^{15}$ to 1$\times$10$^{18}$ cm$^{-2}$, the range in particle density is from 5$\times$10$^{3}$ to 3$\times$10$^{8}$ cm$^{-3}$, the range in temperature is from 25 to 2000 K.

It is impossible to fit all the observed CO lines from \jup= 3 to \jup=22 with a single gas component, while a two-component model gives a good fit to the data. In order to reduce the free parameters we fixed all the parameters that are well constrained by the observations. In particular, the size of the two components can be measured from the maps: the low \jup\, component is extended, that is, we assume a filling factor of 1, and the size of the high \jup\, transitions is measured from the PACS maps to be 18\arcsec. The CO column density of the low \jup\, lines is well constrained from the fitting of the APEX lines at $N$(CO) $\sim$ 6$\times$10$^{16}$ \cmdue, a value consistent with that derived from the rotational diagram. If we fix these three parameters the fitting with the LVG grid of models constrains the other model parameters, as given in Table \ref{tab_lvg}. The first gas component fitting the low \jup\, lines is extended (size$>$ 24\arcsec), cold ($T\sim$80~K) and dense ($n$(H$_2$) = 3$\times$10$^5$ -- 4$\times$ 10$^6$~\cmtre); the second component fitting the high \jup\, lines is compact (size=18\arcsec), warm ($T$ = 1700 -- 2200~K) and with a slightly lower density ($n$(H$_2$) = 2$\times$10$^4$ -- 6$\times$10$^4$~\cmtre). The comparison between the observed CO line fluxes and the best fit model is shown in Fig. \ref{fig_lvg_2comp}. The temperature of the warm gas derived from CO is in agreement with the temperature derived from the H$_2$ lines \citep{giannini11} confirming that high \jup\, CO lines and rotational H$_2$ lines trace the same gas component as also suggested from the spatial coincidence of their emission (Fig.~\ref{fig_zoom_red}). The column density and temperature of the first colder component is in agreement with the estimates derived in the previous section with the rotational diagram, indicating that it is not far from LTE. The temperature is also compatible with the lower limits of the kinetic temperature of the lower excitation gas derived from low \jup\, CO lines in other positions along the outflow \citep{parise06}.

\section{Comparison with shock models}
\label{sect_shock_mod}

In this section, we present comparisons of our observations with outputs of the Paris-Durham shock code \citep{flower15} aimed at constraining the physical and chemical conditions of the two brightest shock positions: knots A and D. To this purpose we used a large dataset of emission lines: we started from H$_2$ {\it Spitzer}-IRS observations \citep{neufeld09,giannini11} and CO lines from APEX \citep{gusdorf15} and our new {\it Herschel}-PACS and HIFI observations, gradually including the new {\it Herschel}-PACS lines from other species: \oi, OH, and H$_2$O. The comparison method requires the combination of a shock model with a radiative transfer code based on the LVG approximation and was originally introduced in \citet{gusdorf08}. Because the datasets available differ in knot A and D, and also because of the different shock structures, we used a slightly different comparison method for each position. 

\begin{figure}
\centering
\includegraphics[width=6.0cm]{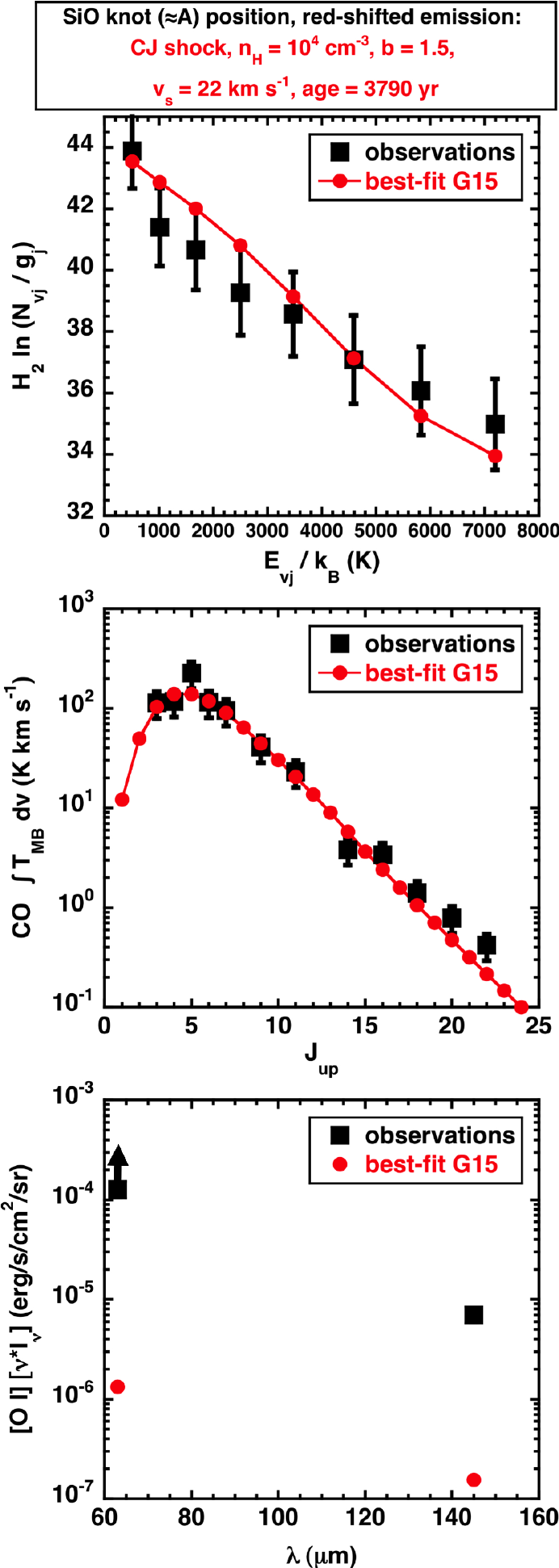}
\caption{Comparison between the best-fit model (red points) obtained in \citet{gusdorf15} from comparisons of a grid of models from the Paris-Durham shock code with observations (black squares) of H$_2$ (from the {\it Spitzer}-IRS telescope) and CO (from the APEX telescope). The target position is the northern shock called SiO-knot in \citet{gusdorf15} and roughly corresponding to knot A of the current study. Upper panel: H$_2$ excitation diagram. Middle panel: CO integrated intensity diagram. Lower panel: \oi\, line flux. The newly presented CO (5--4, 9--8, 14--13, 18--17, 20--19, and 22--21) transitions observed with {\it Herschel} HIFI and PACS have been added to the previously fitted APEX CO lines in the intensity diagram and are satisfactorily fitted by the model.}
\label{bhr71-agfig4}
\end{figure}

\subsection{Knot A}

The shock structure in knot A was already analysed in \citet{gusdorf11} and \citet{gusdorf15}. The observational dataset consisted of lines from H$_2$ 0--0 S($i$) (with $i$ = 0 to 7), CO (\jup = 3, 4, 6, 7, 11, 16) and SiO (\jup = 5, 6, 8). More specifically, 
the observable associated with each molecule was an excitation diagram for H$_2$, and an integrated intensity diagram for CO. These were both used to constrain the input parameters of the shock models. The integrated intensity diagram of SiO was used in a second step to fine-tune our constraints on the pre-shock relative distribution of silicon in the mantle and core of interstellar grains. The CO and SiO spectra pointed towards the existence of red-shifted gas, consistent with a single shock structure. In this region, one non-stationary shock model was consequently found to reproduce observations of H$_2$, CO and SiO, with the following input parameters: pre-shock density of $n$(H) = 10$^4$\cmtre, shock velocity of $\varv_{\rm s}$ = 22~\kms, magnetic field parameter $b$ = 1.5 (defined by $B$($\mu$G) = $b \times$ [$n$(H)(\cmtre)]$^{1/2}$, where $B$ is the intensity of the magnetic field transverse to the shock direction of propagation), and age of $\sim$ 3800 years. 

The first step of our new analysis in knot A was to add the integrated intensities of newly presented CO (5--4, 9--8, 14--13, 18--17, 20--19, and 22--21) transitions observed with {\it Herschel} HIFI and PACS to the CO modelled in \citet{gusdorf15}. We then repeated the analysis done in \citet{gusdorf15} by comparing the CO and H$_2$ observables to outputs from the same grid of models and calculating a chi-square. We found that the best-fit model from \citet{gusdorf15} still provides the best fit to this more complete dataset. The corresponding results for CO and H$_2$ are shown in Fig.~\ref{bhr71-agfig4}.

As we also dispose of \oi\, observations by {\it Herschel}-PACS ($^3P_{\rm 1}$--$^3P_{\rm 2}$ at $\sim$63~$\mu$m and $^3P_{\rm 0}$--$^3P_{\rm 1}$ at $\sim$145~$\mu$m) and we have compared their emissivities to the predictions of the best-fit model found fitting the CO and H$_2$ emission. We did not include the \oi\, lines in the best fitting process because contrary to CO lines they are not velocity-resolved and contrary to H$_2$ they might be subject to self-absorption (especially the lower-lying one, see e.g.~\citealt{leurini15} for a discussion based on velocity-resolved observations in a different region). In particular, the fact that self-absorption might affect the lower lying transition justifies that the corresponding flux should be considered as a lower limit. An additional reason to justify why we did only a posteriori comparison between models and observations of \oi\, lines is the relatively simple assumptions under which their emissivities are modelled: optically thin and in LTE conditions. While it is difficult to diagnose the first optical thickness condition, an indication for the LTE assumption is given by the critical density of the \oi\, 63 \um, of the order of 10$^6$~\cmtre, likely difficult to fulfil in the analysed shock region. 
The result of the comparison of observed \oi\, lines with the best fit model can be seen in the lower panel of Fig.~\ref{bhr71-agfig4}. For both lines, we found a huge discrepancy between observations and models. Indeed the modelled fluxes are 1\% and 2\% of the observed values for the 63~\um\, and 145~\um\, line, respectively. Two explanations can be advanced for this discrepancy: either (i) there is a single shock structure in the observed beam, and the physical ingredients of the Paris-Durham model are not appropriate to describe the physics and chemistry at work in this region, or (ii) a second shock structure exists, strongly emitting in \oi\, and possibly negligibly in CO or H$_2$, such as an atomic jet component. In the first case (i), for example, an enhancement of \oi\, and OH can be the consequence of the photo-dissociation of H$_2$O. The photo-dissociation of molecules can be due to the presence of an UV radiation field that is not taken into account in the version of the Paris-Durham model that we used. However, we would expect this effect to have also a strong impact on the predicted CO and H$_2$ line emission (namely significantly reducing this emission, as these molecules would also undergo dissociation to some extent). Moreover knot A is about 1.3$\times$10$^4$ AU away from IRS1 making the possible proto-stellar radiation field practically non influential at this position.
This is why the second assumption (ii) is a more convincing scenario. Indeed, it is possible that an atomic jet component exists, that would simultaneously be responsible for the \oi\, emission and would not modify the H$_2$ and CO comparisons that have been presented in \citet{gusdorf15} and here. For instance, a dissociative J-type shock propagating in a less dense medium could generate the observed amounts of \oi\, emission, bring a negligible addition in terms of CO line emission, and a weak one in terms of H$_2$ emission. This would be the case, for example, of the stationary model with $n$(H) = 10$^3$~\cmtre, $b = 1.5$, and $\varv_{\rm s}$ = 30--40 \kms\, presented in \citet{hollenbach89}. In such models, the dissociation of molecules is generated by the shock itself. Alternatively, a purely atomic component could be considered, for which models are lacking because of the difficulty to include the presence of self-generated or externally-generated radiation field.
A situation similar to what we observe in knot A of the BHR71 outflow has been observed also in other shocked clumps of the L1157 and L1148 outflows \citep{benedettini12,santangelo13} where it was found that the bright emission of the \oi\, lines cannot be reproduced by only a C-type shock or a low-velocity, non-dissociative J-type shock but requires the presence of an additional dissociative shock that could be associated to the primary jet. Additional evidence of the existence of a fast driving jet were found by {\it Herschel} near the protostar where high velocity gas components were detected, likely produced by a collimated jet very close to the ejecting protostar \citep{kristensen13, nisini15}. 
More observations, with a better spatial and/or spectral resolution, in particular of atomic lines such as \oi, are necessary to unveil the primary jet in outflows and to confirm our current interpretation of \oi\, and OH emission in the shocked spots along the lobes.

Finally, we also compared the prediction of the best-fit model to our {\it Herschel}-PACS observations of OH (lines at 119.2 \um\, and 119.4 \um) and H$_2$O (lines at 108 \um\, and 179 \um). We note that the OH lines and the o-H$_2$O 179 \um\, line might suffer from self-absorption, hence the observed values should be treated as lower limits \citep[e.g.][]{leurini15}. We found that the best-fit model significantly underestimates the integrated intensity of the OH 119.2 \um\, and 119.4 \um\, lines (models: 1.0$\times 10^{-3}$ and 1.8$\times 10^{-3}$~K~km~s$^{-1}$, observations: 0.25 and 0.36~K~km~s$^{-1}$, respectively), similarly as what we obtained for \oi. Conversely, the 108 \um\, and 179 \um\, water lines comparisons provided a much smaller difference: 2.5~K~\kms\, and 41.0~K~\kms\, respectively for the models, and 1.2~K~\kms\, and 8.4~K~\kms\, for the observations. We might be able to further decrease the discrepancy between observations and models for water comparisons by simultaneously fine-tuning the physical and chemical networks in our models, and by using more, preferably velocity-resolved observations of H$_2$O lines. This will be the subject of a dedicated work in a forthcoming publication.

\subsection{Knot D}

\begin{figure}
\centering
\includegraphics[width=8cm]{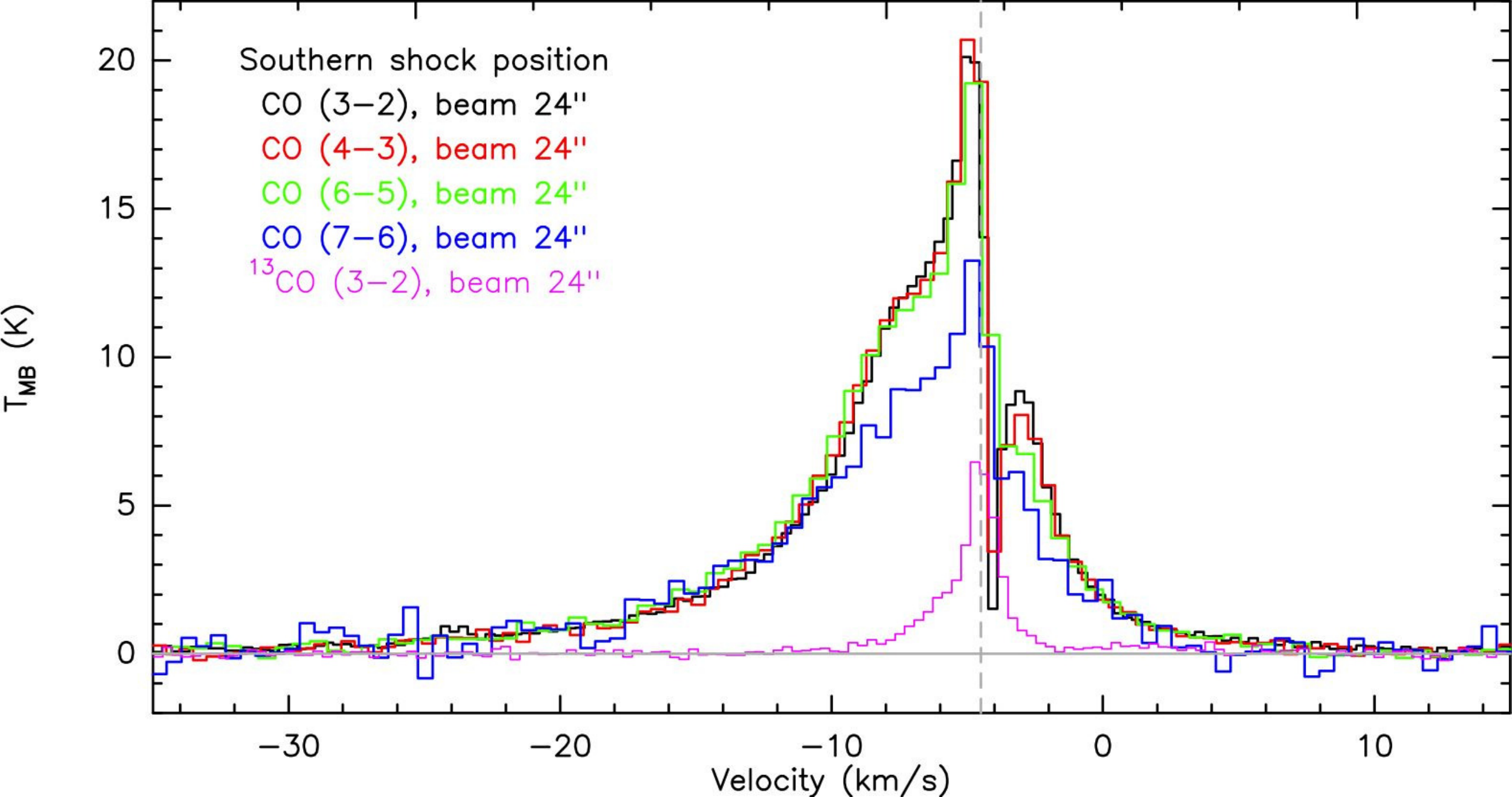}
\caption{CO ladder obtained in knot D with the APEX telescope by \citet{gusdorf15}, convolved to the 24\arcsec\, resolution. The nominal spectral resolution was given in \citet{gusdorf15}.}
\label{bhr71-agfig2}
\end{figure}

\begin{figure}
\centering
\includegraphics[width=8cm]{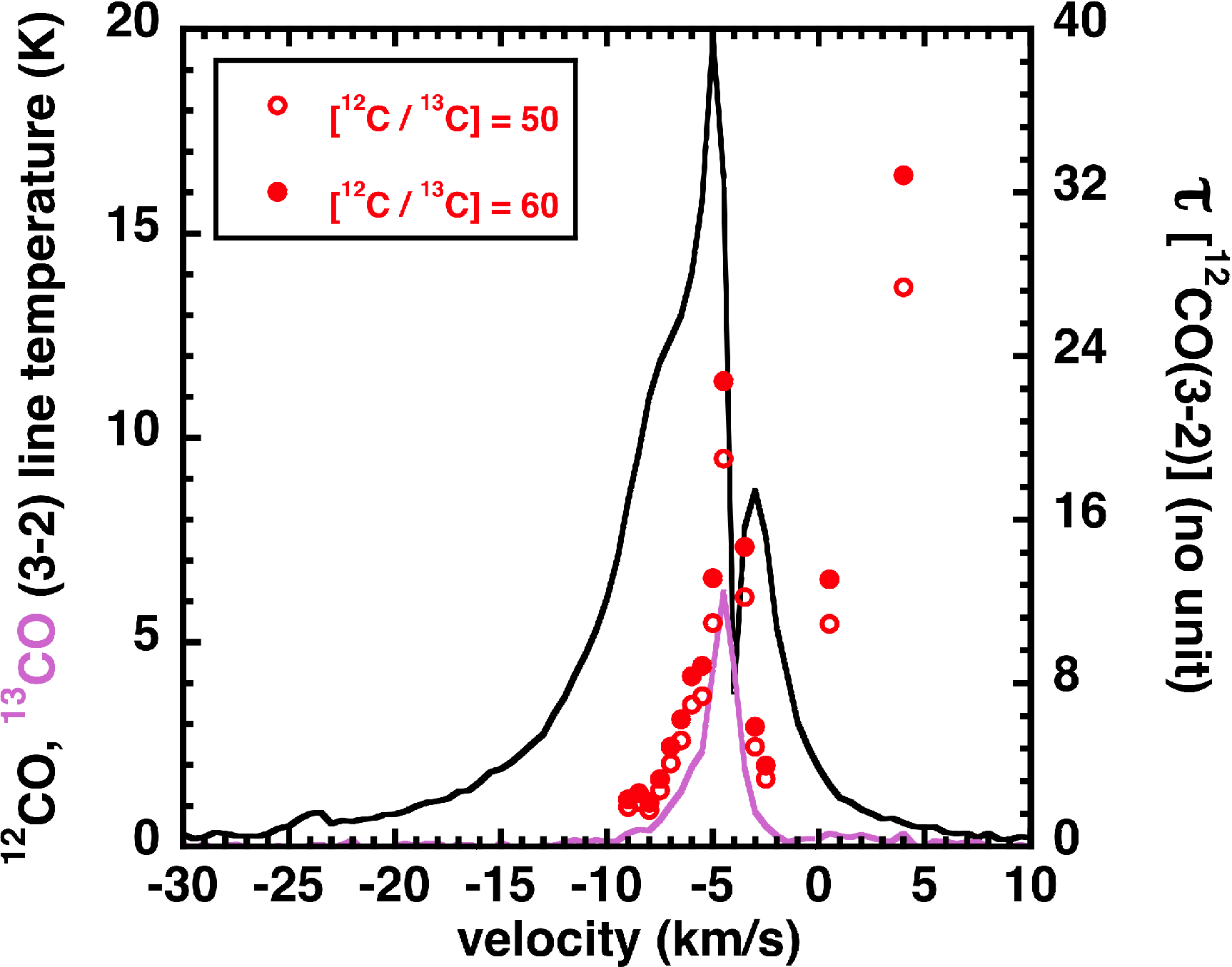}
\caption{Channel-by-channel optical depth of the CO (3--2) line (right-hand y-axis), obtained under two assumption for the [$^{12}$CO]/[$^{13}$CO] relative abundance: 50 and 60 (empty and full red circles), for all channels with at least a 3~$\sigma$ detection in $^{13}$CO (3--2). Also shown the $^{12}$CO and $^{13}$CO (3--2) spectra (in black and pink respectively, left-hand y-axis), obtained in knot D with the APEX telescope.}
\label{bhr71-agfig3}
\end{figure}

The analysis of knot D is slightly different. Indeed, in this region, the southern, blue-shifted lobe of the outflow driven by IRS1 coexists with the southern, red-shifted lobe of the outflow driven by IRS2 (see Fig.~\ref{fig_maps_all}). This can also be seen in Fig.~\ref{bhr71-agfig2}, that shows all spectrally resolved spectra obtained with the APEX telescope \citep{gusdorf15} towards knot D, convolved to a spatial resolution of 24\arcsec\, (for consistency reasons with the analysis in knot A). All spectra indicate the presence of blue-shifted gas up to about -35~\kms\, and of red-shifted gas to about 15 \kms. 
Also shown in Fig.~\ref{bhr71-agfig2} is the APEX $^{13}$CO (3--2) line \citet{gusdorf15}. We used this line in combination with $^{12}$CO (3--2) to estimate the optical thickness of the latter line. Fig.~\ref{bhr71-agfig3} shows spectra of the $^{13}$CO and $^{12}$CO (3--2) lines at the knot D position at the same spectral and spatial resolutions. These spectra allow us to calculate the opacity for each velocity channel  where the $^{13}$CO is detected at more than 3$\sigma$. Assuming an identical excitation temperature for the $^{12}$CO and $^{13}$CO lines, and a typical interstellar abundance ratio of 50--60 \citep[e.g.][]{langer93}, we conclude that the $^{12}$CO (3--2) emission is optically thick at least in the low-velocity regime of the spectral wings. The large optical thickness is consistent with the constancy of the line integrated intensities (in the non-absorbed components) from CO (3--2) and (4--3), when convolved to the same resolution. 

\begin{figure*}
\centering
\includegraphics[width=17cm]{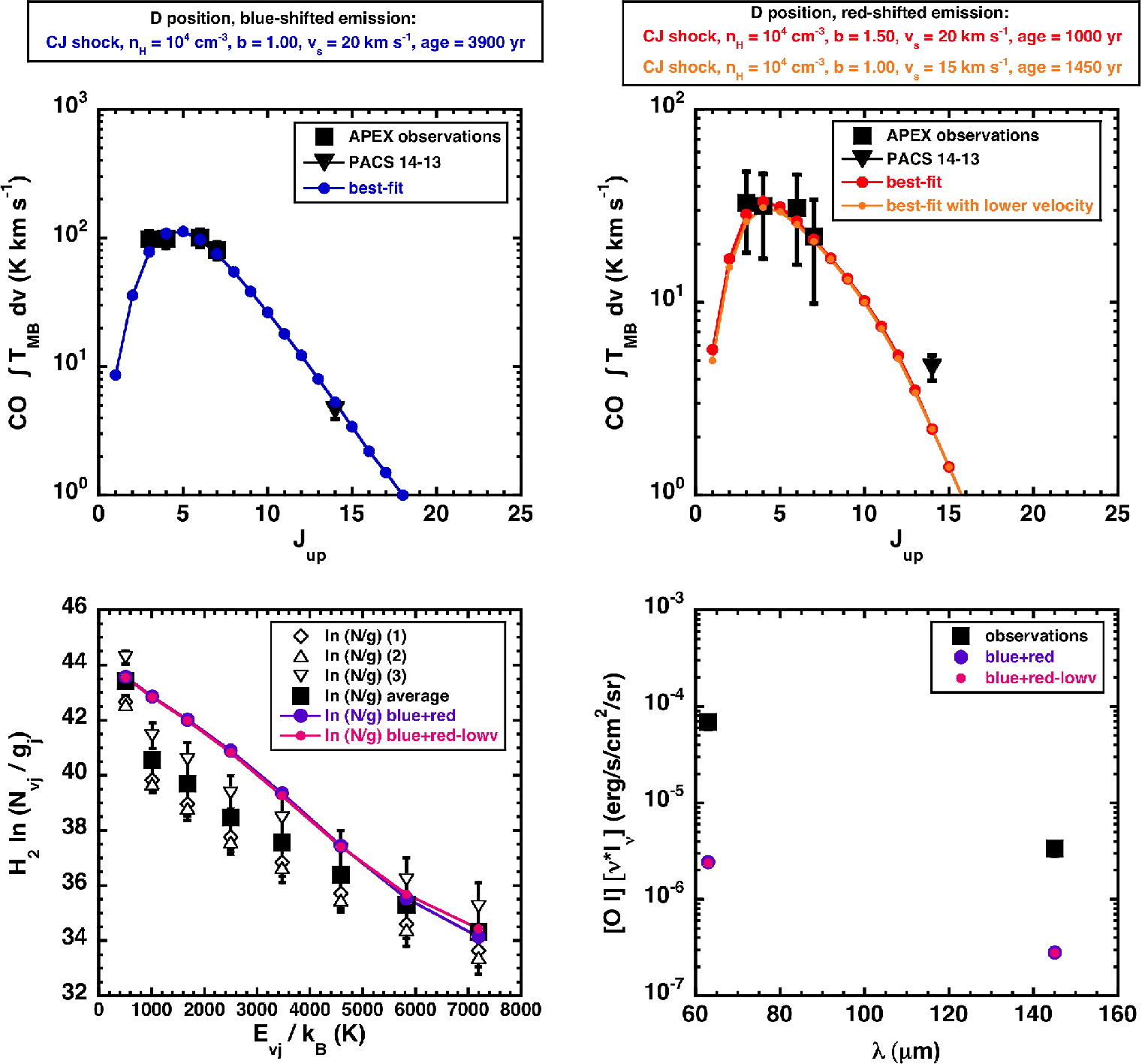}
\caption{Comparison between the best-fit model (coloured points) from comparisons of a grid of models from the Paris-Durham shock code with observations (black squares) of H$_2$ (from the {\it Spitzer}-IRS telescope) and CO (from the APEX telescope). The target position is the southern knot D shock where the blue lobe of the outflow driven by IRS1 overlaps with that driven by IRS2. We consequently fitted two shock layers in this position: one on the red-shifted, one on the blue-shifted gas component. Upper panel, left: CO integrated intensity diagram for the blue component. Upper panel, right: CO integrated intensity diagram for the red component. Lower panel, left: H$_2$ excitation diagram, the three empty symbols correspond to fluxes extracted with different methods (see text for details), the filled squares indicates the average values. Lower panel, right: \oi\, line flux.}
\label{bhr71-agfig5}
\end{figure*}

The presence of blue- and red-shifted gas in knot D compelled us to vary the modelling method with respect to the analysis performed in knot A. Indeed, instead of trying to fit only one shock layer onto the observations, we attempted to model each blue and red lobe independently, using an approach that was already used in \citet{gusdorf12} and \citet{anderl14}. We hence generated two integrated intensity diagrams for CO by using velocity-resolved line spectra (available for \jup = 3, 4, 6, 7): one for the blue-shifted component (from -35~\kms\, to -4.5~\kms), and one for the red-shifted component (from -4.5~\kms\, to 20~\kms). We then compared each of these excitation diagrams to our grid of models to select the best-fit models for knot D. Before doing this, we first extended our grid of models to a larger parameters space.

Indeed, as can be seen in Fig.~\ref{fig_full_sistem}, the position of knot D lies at the edge of the parental Bok globule of the outflow. For this reason, we ran additional shock models (stationary and non-stationary ones) for a lower pre-shock density value: 10$^3$\cmtre. Additionally, we noticed that the linewidth of the red-shifted gas component was smaller than that of the blue-shifted one. We consequently extended our grid towards lower values of the shock velocity (towards 15~\kms included). 

Comparing the velocity-resolved CO lines emission with the extended grids of models, we found that no model with a lower pre-shock density is be able to fit either the blue- or red-shifted gas observations. This is because non-stationary, young models at this pre-shock density are J-type shocks, which are too narrow to generate significant levels of CO emission. Physically, the fact that shocks at the edge of the cloud still propagate in a relatively dense medium can be explained by their propagation in a medium that has already been shock-processed, hence densified by, for example, the primary ejection episodes of the protostar(s). 

The best-fit models for each spectral range are shown in the top panels of Fig.~\ref{bhr71-agfig5}. The input parameters of the best-fit models are: $n$(H) = 10$^4$~cm$^{-3}$, $b$ = 1.0, and $\varv_{\rm s}$ = 20~\kms, and an age of about 3900 years for the blue-shifted CO emission, and $n$(H) = 10$^4$~cm$^{-3}$, $b$ = 1.5, and $\varv_{\rm s}$ = 20~\kms, and an age of about 1000 years for the red-shifted CO emission. For both spectral ranges, the best-fit models are non stationary, with pre-shock density and the $b$ value similar to the values found in knot A. For the red-shifted gas component, we chose to display in Fig.~\ref{bhr71-agfig5} two solutions: the absolute best-fit, with a velocity of 20~\kms\, (red symbols and lines), and another good solution with a velocity closer to the relatively small observed linewidth (15~\kms, orange symbols and lines). The difference between the two solutions is not significant. We remind the reader that when comparing 1D shock models with CO observations, one should not necessarily expect that the observed linewidth should be equal to the shock velocity. Indeed it is possible that shocks propagate in an already moving medium, or are orientated towards the plane of the sky resulting respectively in a lower or higher shock velocity.

Finally, we found an evolutionary trend in the best-fit solution: the best-fit for the blue-shifted emission has an age comparable to what was found for the red-shifted gas in knot A, around 3900 years. Looking at the relative position of the two knots A and D with respect to the exciting source IRS1, one would expect an older age for the more distant knot A than knot D. However, both lines profiles and the best fit shock model suggest that the shock velocity in the southern, blue-shifted lobe should be lower than that in the northern red-shifted lobe. Moreover, the age estimate, as well as all the other input parameters of our shock model, suffer of an uncertainty given by the imperfect coverage of the input parameters space. Therefore, despite the different distance of A and D form the protostar, their age similarity is not too surprising even because both  structures belong to the same outflow.

On the other hand, the age associated with the red-shifted gas (from the outflow driven by IRS2) is younger, 1000 -- 1500 years old. This value is consistent with the dynamical age of this outflow that we estimated from the CO (3--2) map. Indeed, for the outflow driven by IRS2, we found CO (3--2) emission with 10~\kms\, linewidth up to $\sim$56\arcsec\, from the driving source in the northern lobe and $\sim$113\arcsec\, in the southern lobe, translating in a dynamical age comprised between 5000 and 11000 years (also slightly younger than the dynamical age inferred by \citet{gusdorf15} for the outflow driven by IRS1). The uncertainty associated to these values is important, however both the observational dynamical ages and results from models seem to suggest a younger evolutionary stage for IRS2. This is consistent also with the findings of \citet{chen08} based on the analysis of the spectral energy distribution of both IRS1 and IRS2 and with our results shown in Sect. \ref{sect_line_ratio}.

In order to validate these results, we also plotted the integrated intensity of the CO (14--13) line inferred from our {\it Herschel}-PACS measurements. As this observation was not velocity resolved, this value should not be compared to the blue- or red-shifted best-fit, but to their combination. We found that this value is indeed consistent (within the errorbars) with the sum of the integrated intensity of the two best-fit. Similarly, we used the H$_2$ observations by {\it Spitzer}-IRS to further check out model results. We extracted the level populations in various methods already discussed in \citet{gusdorf15}, resulting in the set of symbols shown in the lower left-hand panel of Fig.~\ref{bhr71-agfig5}. Since these observations are not velocity resolved, we summed the contribution from the two best-fits and compared them with the observed excitation diagram. The result can be seen in colours in the lower left-hand panel of Fig.~\ref{bhr71-agfig5}. Two sets of models are plotted, since we chose to display the sum of the two best-fit contributions (purple symbols and line) and the sum of the contributions from the blue-shifted best-fit with that from the red-shifted best-fit with a low velocity (pink symbols and line). In both cases, the results are fitted with a similar quality as in knot A. We note that the fit is not perfect, and we are working on solutions to improve our models, specially from the point of view of the geometrical description of the observed region. Similarly, we summed either the \oi\, fluxes from our two best-fit models or the blue-shifted gas best-fit with the red-shifted gas best fit with the lower velocity and compared these values to the observations. The results can be seen in the right-hand lower panel of Fig.~\ref{bhr71-agfig5}. Similarly to what found for knot A, also for knot D the best fit models drastically underestimate the emission of both \oi\, transitions. We refer to the previous subsection for an explanation of this result. 

Finally, in this knot, we have observed H$_2$O emission from the lines at 108 \um\, and 179 \um\, with PACS, that is, without spectrally resolving the line profile. We could then compare the values predicted by the sums of our best models: 39.8 -- 45.2~K~km~s$^{-1}$ and 2.5 -- 2.9~K~km~s$^{-1}$, respectively with the observed integrated intensities: 11.7~K~km~s$^{-1}$ and 1.0~K~km~s$^{-1}$. Given the possible self-absorption (see discussion in the above subsection), the observed values are lower limits. We note that a discrepancy between models and observations does exist, but less impressive than for the \oi\, lines, similarly as the conclusions reached in the previous paragraph for knot A, with the same conclusions.

\section{Summary and conclusions}

We presented {\it Herschel} maps in \coquattordoci, \watcsn, \watco\, and \oi\, 145 \um\, of the BHR 71 outflows system plus additional CO lines measurements towards two outflow positions.
These far infrared maps show seven bright knots surrounded by a low level extended emission. Two knots correspond to the two protostars IRS1 and IRS2 exciting the two outflows composing the BHR 71 system while the other knots are located in the outflow lobes and represent shock episodes along the outflow jet. Indeed, the far infrared emission peaks, spatially coincident in all the observed PACS lines, are roughly coincident with the positions of the local maxima of H$_2$ lines and of the fastest outflowing gas traced by the \cosei\, line. 
In the southern lobe the blue-shifted gas of the IRS1 outflow and the red-shifted gas of the IRS2 outflow partially overlap, however several evidences indicate that the emission in the observed PACS lines is dominated by the flux emitted from gas associated with the blue lobe of the IRS1 outflow.

The ratio of the {\it Herschel}-PACS lines intensity, between lines of the same molecular species, have been calculated in the seven knots. The ratios are quite similar within the errors, showing that the excitation conditions of the fast moving gas do not change significantly along the outflow, apart a lower density and temperature at the extremity of the southern blue lobe where the outflow is flowing outside the parental molecular cloud. More peculiar ratios are found towards the two protostars from which we deduced that IRS1 have a much higher temperature ($T>$ 1200 K) and that a significant extinction is present towards IRS2 able to absorb part of the \oi\, 63\um\, emission. 

In the brightest position of the red lobe of the IRS1 outflow (knot A) we observed additional CO lines with \jup\, up to 22. Combining our {\it Herschel} observations with measurements of the low \jup\, CO, H$_2$ and SiO lines \citep{gusdorf11,gusdorf15,neufeld09} we were able to make a deep analysis of the physical and shock conditions present in this pure shock position. Rotational diagram, spectral profile shape and LVG analysis of the CO lines showed the presence of two gas components: one extended (size$>$ 24\arcsec), cold ($T\sim$80~K) and dense ($n$(H$_2$) = 3$\times$10$^5$ -- 4$\times$10$^6$~\cmtre) and another compact (18\arcsec), warm ($T$ = 1700 -- 2200~K) and slightly less dense ($n$(H$_2$) = 2$\times$10$^4$ -- 6$\times$10$^4$~\cmtre). The CO column density of the cold component is about one order of magnitude higher than that of the warm component.

In the two infrared brightest knots, knot A in the northern lobe and knot D in the southern lobe, we compared observations with the Paris-Durham shock code. We found that non-stationary shock models well reproduce observations of CO and H$_2$ in both knots. In knot A, the best-fit model is non stationary with $n$(H) = 10$^4$~cm$^{-3}$, $b$ = 1.5, and $\varv_{\rm s}$ = 22~\kms, and an age of about 3800 years. In knot D, two shocks coexist the blue-shifted lobe of the IRS1 outflow and the red-shifted lobe of the IRS2 outflow. The blue-shifted CO emission can be described well with a non-stationary shock with $n$(H) = 10$^4$~cm$^{-3}$, $b$ = 1.0, and $\varv_{\rm s}$ = 20~\kms, and an age of about 3900 years, conditions very similar to that found in the red-shifted counterpart. On the other hand, the red-shifted CO emission from the IRS2 outflow can be reproduced by a non-stationary shock with similar parameter, $n$(H) = 10$^4$~cm$^{-3}$, $b$ = 1.5, and $\varv_{\rm s}$ = 20~\kms, but a younger age of about 1000 years. Because of the smaller linewidth associated with this component, we also investigated the possibility to fit the observations with a lower velocity shock with $n$(H) = 10$^4$~cm$^{-3}$, $b$ = 1.0, and $\varv_{\rm s}$ = 15~\kms, and an age of about 1450 years. Interestingly, the younger age of the outflow powered by IRS2 is in agreement with the evolutionary estimate derived from the spectral energy distribution fitting of the two driving protostars \citep{chen08} and with the higher obscuration of the IRS2 protostar deduced from the \oi\, lines ratio.

The non-stationary shock models that  well reproduce the CO and H$_2$ observations underestimate by about two orders of magnitude the observed \oi\, lines at 63 \um\, and 145 \um, and the OH lines at 119.2 \um\, and 119.4 $\mu$m, when available. We interpret this discrepancy as the evidence for the existence of a primary, mostly atomic, jet component in the outflow. In fact a shock with a radiative precursor, with $n$(H) = 10$^3$~cm$^{-3}$, $b$ = 1.5, and $\varv_{\rm s}$ = 30--40~\kms\, as presented in \citet{hollenbach89} can account for the observed \oi\, emission without significantly altering our conclusions with respect to CO and H$_2$. Our {\it Herschel}-PACS observations of \oi\, and OH, however, are both spatially and spectroscopically unresolved, therefore they do not directly probe the jet. Moreover, our shock model does not account for the effects of a possible UV radiation field affecting the shock (whether this UV field is created by the shock itself or emanates from the protostar). It is worth noting that several other {\it Herschel} observations both close to protostars and in the outflow lobes have been interpreted as evidence of dissociative shocks produced by the primary jet (\citealt{benedettini12, kristensen13, nisini15}) but only additional observations of species such as OH, [\ion{C}{ii}] and \oi, possibly spectroscopically resolved, for example with SOFIA, could directly probe the primary collimated jet in outflows.

Finally, we compared the predictions of the models that best fit the CO and H$_2$ emission in knots A and D to water observations. We found small discrepancies that might be resolved by further investigations on the shock modelling of water, and their applications to a larger number of velocity-resolved observations of water lines. This will be the object of a forthcoming publication.

Our results in the BHR71 outflow are consistent with other recent studies of outflows that have carried out global analysis of emission lines with different excitation temperatures, for example the CO lines with low \jup\, observable from the ground together with the intermediate and high \jup\, lines observed with {\it Herschel} (e.g. \citealt{lefloch12, benedettini12}) or large samples of H$_2$O lines observed with {\it Herschel} (e.g. \citealt{santangelo13,busquet14}). These studies have shown the presence of multiple components in the molecular outflowing gas at different temperatures and densities. So far, three components have been detected in most outflows and have been associated by the authors to the following physical origin: a cold component with temperature of about 60 - 80 K associated to the walls of the cavity formed by the entrained gas, a warm component with temperature of about 200 - 500 K associated to a non-dissociative shock and a hot component with temperature of about 800 - 2000 K, also emitting in H$_2$ rotational lines and atomic lines, associated with a dissociative J-type shock in the compact region where the protostellar jet impacts the less fast gas. In this respect we found that the chemically active BHR71 outflow, as well as L1157, the other well studied chemically active outflow \citep{benedettini12,lefloch12,busquet14}, does not show evident peculiarity with respect to less chemically active outflows. This is also true when comparing the luminosities of the main far-infrared coolants and the mass loss rate that instead are correlated to the bolometeric luminosity and mass of the envelope of the exciting protostar \citep{neufeld09,karska13, tafalla13, nisini15}.

The results presented in this paper show that the BHR 71 outflow is an ideal astrophysical laboratory for studying shocks in protostellar outflows. However, the major limitation of the analysis performed on the large dataset already acquired for this source is the exiguous spatial resolution of the single dish observations.  Improving the spatial resolution with interferometric data will allow us to make a step forward in modelling the physics and the chemistry of protostellar shocks. This makes the BHR 71 outflow one of the best candidates for future ALMA observations.

\begin{acknowledgements}
We thank Prof. D. Neufeld for providing us with the reduced Spitzer data. 
GB acknowledges the support of the Spanish Ministerio de Economia y Competitividad (MINECO) under the grant FPDI-2012-18204. GB is supported by the Spanish MICINN grant AYA2014-57369-C3-1-P.
PACS has been developed by a consortium of institutes led by MPE (Germany) and including UVIE (Austria); KU Leuven, CSL, IMEC (Belgium); CEA, LAM (France); MPIA (Germany); INAF-IFSI/OAA/OAP/OAT, LENS, SISSA (Italy); IAC (Spain). This development has been supported by the funding agencies BMVIT (Austria), ESA-PRODEX (Belgium), CEA/CNES (France), DLR (Germany), ASI/INAF (Italy), and CICYT/MCYT (Spain).
HIFI has been designed and built by a consortium of institutes and university departments from across Europe, Canada and the United States under the leadership of SRON Netherlands Institute for Space Research, Groningen, The Netherlands and with major contributions from Germany, France and the US. Consortium members are: Canada: CSA, U.Waterloo; France: CESR, LAB, LERMA, IRAM; Germany: KOSMA, MPIfR, MPS; Ireland, NUI Maynooth; Italy: ASI, IFSI-INAF, Osservatorio Astrofisico di Arcetri-INAF; Netherlands: SRON, TUD; Poland: CAMK, CBK; Spain: Observatorio Astronómico Nacional (IGN), Centro de Astrobiología (CSIC-INTA). Sweden: Chalmers University of Technology - MC2, RSS \& GARD; Onsala Space Observatory; Swedish National Space Board, Stockholm University - Stockholm Observatory; Switzerland: ETH Zurich, FHNW; USA: Caltech, JPL, NHSC. 
The Digitized Sky Surveys were produced at the Space Telescope Science Institute under U.S. Government grant NAG W-2166. The images of these surveys are based on photographic data obtained using the Oschin Schmidt Telescope on Palomar Mountain and the UK Schmidt Telescope. The plates were processed into the present compressed digital form with the permission of these institutions. The digitized images of the southern sky are copyright \textcopyright 1993-5 by the Anglo-Australian Observatory Board, and are distributed herein by agreement. 
\end{acknowledgements}

\end{document}